\documentclass[12pt,preprint]{aastex}
\usepackage{emulateapj5}

\slugcomment{To appear in The Astrophysical Journal}

\shorttitle{Lyman Break Galaxies at $z\simeq 4$ and $5$}
\shortauthors{Ouchi et al.}

\begin{document}


\title{
Subaru Deep Survey VI.\\
A Census of Lyman Break Galaxies at $z\simeq 4$ and $5$\\
in the Subaru Deep Fields: Clustering Properties\altaffilmark{1}
}


\author{Masami Ouchi        \altaffilmark{2},
        Kazuhiro Shimasaku  \altaffilmark{2,3},
        Sadanori Okamura    \altaffilmark{2,3},\\
	Hisanori Furusawa   \altaffilmark{4},
	Nobunari Kashikawa  \altaffilmark{5},
	Kazuaki Ota         \altaffilmark{5},\\
        Mamoru Doi          \altaffilmark{6},
        Masaru Hamabe       \altaffilmark{7},
        Masahiko Kimura     \altaffilmark{8},
        Yutaka Komiyama     \altaffilmark{4},\\
        Masayuki Miyazaki   \altaffilmark{2},
        Satoshi Miyazaki    \altaffilmark{4},
        Fumiaki Nakata      \altaffilmark{9},\\
        Maki Sekiguchi      \altaffilmark{10},
        Masafumi Yagi       \altaffilmark{5}, and
        Naoki Yasuda        \altaffilmark{5}
        }

\email{ouchi@astron.s.u-tokyo.ac.jp}


\altaffiltext{1}{Based on data collected at 
        Subaru Telescope, which is operated by 
        the National Astronomical Observatory of Japan.}
\altaffiltext{2}{Department of Astronomy, School of Science,
        University of Tokyo, Tokyo 113-0033, Japan}
\altaffiltext{3}{Research center for the Early Universe, School of Science,
        University of Tokyo, Tokyo 113-0033, Japan}
\altaffiltext{4}{Subaru Telescope, National Astronomical Observatory, 
        650 N.A'ohoku Place, Hilo, HI 96720, USA}
\altaffiltext{5}{National Astronomical Observatory, 
        Mitaka, Tokyo 181-8588, Japan}
\altaffiltext{6}{Institute of Astronomy, School of Science, 
        University of Tokyo, Mitaka, Tokyo 181-0015, Japan}
\altaffiltext{7}{Department of Mathematical and Physical Sciences,
        Faculty of Science, Japan Women's University, Tokyo 112-8681, Japan}
\altaffiltext{8}{Department of Astronomy, Kyoto University, 
        Sakyo-ku, Kyoto 606-8502}
\altaffiltext{9}{Department of Physics, University of Durham,
        South Road, Durham DH1 3LE, UK}
\altaffiltext{10}{Institute for Cosmic Ray Research, 
        University of Tokyo, Kashiwa, Chiba 277-8582}


\begin{abstract}

%
%
We investigate the clustering properties of 
2,600 Lyman Break Galaxies (LBGs) at $z=3.5-5.2$
in two large blank fields,
the Subaru Deep Field and the Subaru/XMM Deep 
Field (600arcmin$^2$ each).
%
%
The angular correlation functions of these LBGs 
show a clear clustering at both $z\simeq 4$ and $5$.
The correlation lengths are 
$r_0= 4.1^{+0.2}_{-0.2}$ and $5.9^{+1.3}_{-1.7} h_{100}^{-1}$ Mpc 
($r_0= 5.1^{+1.0}_{-1.1}$ and $5.9^{+1.3}_{-1.7} h_{100}^{-1}$ Mpc)
for all the detected LBGs (for $L\gtrsim L^*$ LBGs) 
at $z\simeq 4$ and $5$, respectively.
These correlation lengths correspond to 
galaxy-dark matter biases of 
$b_g=
2.9^{+0.1}_{-0.1}$ and
$4.6^{+0.9}_{-1.2}$
($b_g=3.5^{+0.6}_{-0.7}$ and $4.6^{+0.9}_{-1.2}$), 
for all the detected LBGs (for $L\gtrsim L^*$ LBGs) 
at $z\simeq 4$ and $5$, respectively.
These results, 
combined with estimates for $z\simeq 3$ LBGs in the literature, 
show that the correlation length of $L\gtrsim L^*$ LBGs 
is almost constant, $\sim 5 h_{100}^{-1}$ Mpc, 
over $z\simeq 3-5$, 
while the bias monotonically increases with redshift at $z\gtrsim 3$.
We also find that for LBGs at $z\simeq 4$ 
the clustering amplitude increases with UV-continuum luminosity 
and with the amount of dust extinction.
%
%
We estimate the mass of dark halos hosting various kinds of 
high-$z$ galaxies including LBGs 
with the analytic model given by \citet{sheth1999}. 
We find that the typical mass of dark halos 
hosting $L\gtrsim L^*$ LBGs is about
$1\times 10^{12} h_{70}^{-1}M_\odot$ over $z\simeq 3-5$, 
which is comparable to that of the Milky Way Galaxy. 
A single dark halo with 
$\sim 10^{12} h_{70}^{-1} M_\odot$ is found to host 
0.1-0.3 LBG on average but host about four 
$K$-band selected galaxies.

\end{abstract}




\section{Introduction}
\label{sec:introduction}

Formation history of galaxies is basically 
understood by two fundamental evolutionary 
processes, i.e., production of stars 
and accumulation of dark matter. 
Electromagnetic radiation from galaxies gives 
information about the status of baryonic matter, i.e.,
stars, gas, and dust. 
Although dark matter should play a key role in galaxy
formation, it cannot be directly detected using
electromagnetic waves.
Clustering properties of galaxies are closely related to
the distribution and amount of underlying dark matter
(e.g., \citealt{peacock2001,percival2001,verde2002,percival2002,
szalay2003}). 
Thus, measuring clustering properties of galaxies is
very useful for understanding the dark side of galaxy 
properties (different from stellar properties).

Clustering properties in the present-day universe
have been well investigated using recent wide-field
surveys such as 2dFGRS \citep{colless2001}
and SDSS \citep{york2000}.
Clustering properties of galaxies are derived from
spatial or angular two-point correlation functions (CFs), and 
are compared with those of the dark matter predicted 
by Cold Dark Matter (CDM) models \citep{peacock2001,
percival2001,verde2002,percival2002,scranton2002,
connolly2002,dodelson2002,tegmark2002,zehavi2002,szalay2003}.
An analysis of the power spectrum from 2dFGRS data 
shows that the distribution of galaxies agrees with that of
dark matter \citep{verde2002}. In other words,
optically selected galaxies
trace the underlying mass distribution at $z=0$.

Clustering properties of galaxies are now 
investigated at higher redshifts 
up to $z\sim4$. 
%
%
%
\citet{daddi2000} have reported
a strong angular correlation for EROs at $z\sim 1$,
and estimated the correlation length of the spatial CF
to be $r_0=12\pm3 h_{100}^{-1}$ Mpc, which
is comparable to that of present-day ellipticals
\citep{daddi2001}. \citet{mccarthy2001} found 
similarly large, but slightly smaller, correlation lengths
($r_0\simeq 9-10 h_{100}^{-1}$ Mpc) for their red galaxies
at $z\sim 1$.
\citet{miyazaki2003} found strong clustering 
for both old galaxies and dusty star-forming galaxies 
beyond $z \sim 1$.
%
%
%
%
\citet{giavalisco1998} have studied 
CFs of Lyman Break Galaxies (LBGs) 
on the basis of a large sample for bright ($R<25.5$) LBGs 
lying around $z=3$ selected by $U_nGR$ colors.
They have found that the spatial distribution of $z=3$ LBGs
is strongly biased relative to the 
dark matter distribution predicted by CDM
models with a linear bias of 4.5 for
an Einstein-de Sitter cosmology 
(see also \citealt{adelberger1998}).
In a subsequent paper, \citet{giavalisco2001}
have found the clustering amplitude 
of $z\sim 3$ LBGs to depend on their rest-frame 
ultraviolet luminosity,
with fainter galaxies less strongly clustered,
which is a similar property to that 
found in the present-day universe.
\citet{arnouts1999,arnouts2002} have measured the correlation length of 
faint galaxies in the HDF-N and HDF-S
over the redshift range $0<z<4$,
though errors in their measurements are large 
due to small number statistics. 
Recent observational
results show that LBGs at $z\simeq 4$ and Lyman $\alpha$
emitters (LAEs) at $z=4.9$ 
are strongly biased against underlying dark matter, and 
that the clustering amplitudes of LAEs are segregated with 
respect to Ly$\alpha$
luminosity (\citealt{ouchi2001a};
\citealt[hereafter SDS II]{ouchi2003a}).

Narrow-band searches of LAEs on targeted fields have found strong
clustering of LAEs around quasars 
\citep{campos1999,moller2001,stiavelli2001,fynbo2003},
radio galaxies \citep{venemans2002},
high redshift clusters \citep{palunas2000},
and overdensities in the LBG distribution 
\citep{steidel2000}.
\footnote{
Searches of LBGs in targeted fields are also made
by, e.g., \citet{bouche2003} who find a significant
clustering of LBGs around a damped Ly $\alpha$ absorber
at $z=3$.
}
Recent deep and wide-field LAE searches with Subaru/Suprime-Cam
have revealed a large-scale distribution of LAEs (SDS II),
and \citet[hereafter SDS IV]{shimasaku2003}
have found a filamentary
large-scale structure made of LAEs at $z=4.9$
whose width and length are $20 h_{70}^{-1}$ Mpc and $>50 h_{70}^{-1}$ Mpc,
respectively. The size of this high-$z$ large-scale structure
is comparable to those of present-day large-scale structures
found by \citet{geller1989}. The existence of
such a high-$z$ large-scale structure also indicates
that  galaxies at $z\simeq 5$ would be highly biased against
underlying dark matter.
However, LAE searches by narrow-band imaging
observe a thin slice of high-$z$ universes,
resulting in a surveyed volume of as small
as about an order of $10^4 h_{70}^{-3}$ Mpc$^3$. Such surveys may suffer from
a strong field variance. 
On the other hand, survey volumes
for LBGs are generally about 20-50 times larger than those for
LAEs. Thus, measuring clustering properties of
LBGs is highly desirable to quantify the average distribution of
high-$z$ galaxies.

Motivated by this, 
we carried out deep and wide-field imaging for two blank fields,
the Subaru Deep Field
(SDF: $13^h 24^m 21.4^s$,$+27^\circ 29 ' 23''$[J2000];
\citealt[hereafter SDS I]{maihara2001}) 
and the Subaru/XMM Deep survey Field
(SXDF: $2^h 18^m 00^s$,$-5^\circ 00' 00''$[J2000];
Sekiguchi et al. 2003 in preparation; see also \citealt{ouchi2001a}),
and made samples of 2,600 LBGs at $z=3.5-5.2$ distributed in
these two fields (600 arcmin$^2$ each).
Details of the samples are described in
\citet[hereafter SDS V]{ouchi2003d}.

In this paper,
we investigate clustering properties of LBGs at $z=4$ and $5$
using these large LBG samples.
We then compare the observational results
with predictions of CDM models as follows. 
Applying analytic models for the spatial clustering of dark matter
to the observational results,
we estimate the mass of dark halos which host galaxies
from the bias of galaxy-dark matter distribution
\citep{bullock2002,moustakas2002}.
Since the dark-halo mass
is a fundamental quantity of galaxies,
we next compare
various high-$z$ galaxies detected in various wavelengths
in terms of halo mass.
In reality, a large fraction of high-$z$ galaxies detected to date are
either LBGs, LAEs, EROs, or SCUBA sources. 
%
%
These high-$z$ galaxy samples are biased toward UV-bright
star-forming galaxies (LBGs), strong-Ly$\alpha$ emission
galaxies (LAEs), old passive galaxies (EROs), and
dusty starburst galaxies (SCUBA sources).
It is essential to understand the nature of
the high-$z$ galaxy population as a whole
by combining these biased samples
in order to propose a scenario for the evolution of
galaxies at high redshifts.
Finally, 
we examine the relation between
high-$z$ galaxies and present-day galaxies
on the basis of dark-halo mass of descendants
predicted by the CDM model \citep{moustakas2002}.
%

The outline of this paper is as follows. 
In section \ref{sec:data}, 
we describe photometric samples of LBGs
at $z=3.5-5.2$ from the data in the SDF and the SXDF,
which are obtained by SDS V.
These are the largest samples of LBGs at $z\gtrsim 4$
in contiguous areas obtained to date, 
and thus enable detailed studies of clustering properties 
of galaxies at the highest redshifts so far.
Section \ref{sec:spatial} presents 
observational results. We investigate the spatial distributions
of LBGs to obtain the observational results. 
We derive angular correlation
functions (CFs) of LBGs, and calculate correlation
lengths of spatial clustering.
We estimate the galaxy-dark matter bias of LBGs 
at $z\simeq 4$ and $5$, and 
compare them with those of low-$z$ galaxies and LBGs at $z=3$.
We examine the dependence of clustering amplitude
on luminosity and color for LBGs and LAEs.
Section \ref{sec:galaxy} presents theoretical implications.
We compare our observational results
on the CF and the luminosity function obtained in SDS V 
with an analytic CDM model
to investigate properties of
dark halos hosting
high-$z$ galaxies and the relation between
galaxies (luminous matter) and dark halos.
In Section \ref{sec:galaxy}, we also infer the mass of descendants
of these high-$z$ galaxies.
In section \ref{sec:discussions}, 
we propose a unified view for a variety of
high-$z$ galaxies and discuss the formation history 
of galaxies from $z=5$ to $z=0$,
based on the results
obtained in this paper and SDS V.
Section \ref{sec:conclusions} presents a summary of this paper.

Throughout this paper, magnitudes are in the AB system
\citep{oke1974,fukugita1995}.
The values for the cosmological parameters adopted in this paper
are: $(\Omega_m,\Omega_\Lambda,n,\sigma_8)=
(0.3,0.7,1.0,0.9)$. These values are 
the same as those obtained 
from the latest CMB observations
\citep{spergel2003}.
\footnote{
\citet{spergel2003} show that the best-fit parameters
are $n=0.93\pm0.03$ and $\sigma_8=0.84\pm0.04$.
However, the differences between these values and 
those we adopt here are less than 10\%.
}
We adopt $n=1$ and $\sigma_8=0.9$, 
since these values were used
in many of the theoretical and 
observational studies to date.
We express physical quantities using
$h_{70}$, where $h_{70}$ is the Hubble constant
in units of 70 km s$^{-1}$ Mpc$^{-1}$. The exception
is the correlation length, $r_0$. We express $r_0$ using
$h_{100}$, where $h_{100}$ is the Hubble constant
in units of 100 km s$^{-1}$ Mpc$^{-1}$, since most
of the previous studies use $h_{100}$ to express $r_0$.

\section{Observations and Galaxy Samples}
\label{sec:data}

We use three Lyman Break Galaxy (LBG) samples
presented in \citet{ouchi2003d}, i.e.,
SDS V. Details of the samples are
shown in SDS V, and we give a brief summary
of the data and the samples in the following.
In sections \ref{sec:spatial}-\ref{sec:galaxy},
we describe the clustering properties of 
a sample of Lyman $\alpha$ emitters (LAEs) at $z=4.86\pm0.03$
found in the SDF to compare them with
those of LBGs. Details of the LAE sample
and their clustering properties (e.g., distribution 
and angular correlation function etc.) 
are shown in SDS II.

\subsection{Observations}
\label{sec:definition}

During the commissioning runs of Suprime-Cam from
November 2000 to November 2001,
we carried out multi-band, deep and wide-field 
optical imaging in two blank fields.
One is the Subaru Deep Field
(SDF: $13^h 24^m 21.4^s$,$+27^\circ 29 ' 23''$[J2000])
covering a 616 arcmin$^2$ area 
whose central region has very deep $J$ and $K'$ images
(SDS I).
The other is the Subaru/XMM Deep Field
(SXDF: $2^h 18^m 00^s$,$-5^\circ 12 ' 00''$[J2000])
covering a 653 arcmin$^2$ area.
We took multi-color images with the
$B$-, $V$-, $R$-, $i'$-, and $z'$-band filters for both fields.
The exposure time ranges from 
81 to 210 (40 to 177) minutes among the bandpasses 
in the SDF (SXDF).
The limiting magnitude and the seeing size
of the reduced $i'$ image is $i'=26.9$ and $0''.9$ 
($i'=26.2$ and $1''.0$) for the SDF (SXDF), respectively.
We limit the object catalogs 
to $i'<26.5$ and $z'<26.0$ 
($i'<26.0$ and $z'<25.5$)
for the SDF (SXDF),
in order to provide 
a reasonable level
of photometric completeness. Our $i'$-limited and $z'$-limited
catalogs contain respectively 45,923 (39,301) and 37,486 (34,024) 
objects for the SDF (SXDF).

\subsection{Definition of $BRi$, $Viz$, and $Riz$-Lyman Break Galaxies}
We make three photometric samples of LBGs 
by the following two color selections.
The first sample is for $BRi$-LBGs at $z\sim4$.
Their Lyman break enters into the $B$ band and thus
they are identified by red $B-R$ and blue $R-i'$ colors.
Similarly, the second is for $Viz$-LBGs at $z\sim5$
selected by $V-i'$ and $i'-z'$, and the
third is for $Riz$-LBGs at $z\sim5$ selected by $R-i'$ and $i'-z'$. 
We define the selection criteria for these LBGs
so that the completeness be sufficiently high and
the contamination be negligibly small.
In order to determine the selection criteria
in the two-color diagrams, 
we use the best-fit SEDs of galaxies given in 
\citeauthor{furusawa2000}'s (\citeyear{furusawa2000})
HDF-N photometric redshift catalog.
We generate artificial galaxies that mimic the colors of 
the HDF-N galaxies,
and distribute them randomly on our original images
after adding Poisson noises.
Then, we detect these simulated objects and measure their
brightness in the same manner as for our original images.
We iterate this process 100 times, and generate
probability maps of low-$z$ interlopers and 
LBGs in the two-color diagrams.
Based on the probability maps, we determine the selection
criteria for LBGs which give small contaminants and
keep a sufficiently high completeness.
The selection criteria adopted are
{\footnotesize
\begin{eqnarray}
\label{eq:lbgselection_BRiLBG}
B-R>1.2,\ R-i'<0.7,\ B-R>1.6(R-i')+1.9\ \ {\rm ({\it BRi}-LBGs)}\ \ \ \\
\label{eq:lbgselection_VizLBG}
V-i'>1.2,\ i'-z'<0.7,\ V-i'>1.8(i'-z')+1.7\ \ {\rm ({\it Viz}-LBGs)}\ \ \ \\
\label{eq:lbgselection_RizLBG}
R-i'>1.2,\ i'-z'<0.7,\ R-i'>1.0(i'-z')+1.0\ \ {\rm ({\it Riz}-LBGs)}\ \ .
\end{eqnarray}
}

We apply these selection criteria to our photometric catalogs,
and find 1,438 (732), 246(34), and 68 (38) objects for
$BRi$-LBGs, $Viz$-LBGs, and $Riz$-LBGs in the SDF (SXDF).
The $V$-band image of the SXDF is not deep enough 
to produce a $Viz$-LBG sample 
with a small number of contaminants 
by eq.(\ref{eq:lbgselection_VizLBG}).
In order to avoid a high contamination rate,
we adopt
\begin{equation}
V-i'>1.2,\ i'-z'<0.7,\ V-i'>1.8(i'-z')+2.3
\label{eq:lbgselection_SXDFVizLBG}
\end{equation}
as the $Viz$-LBG selection criteria for the SXDF.
Table \ref{tab:sample} summarizes the LBGs found in our data.

\begin{figure}
\plotone{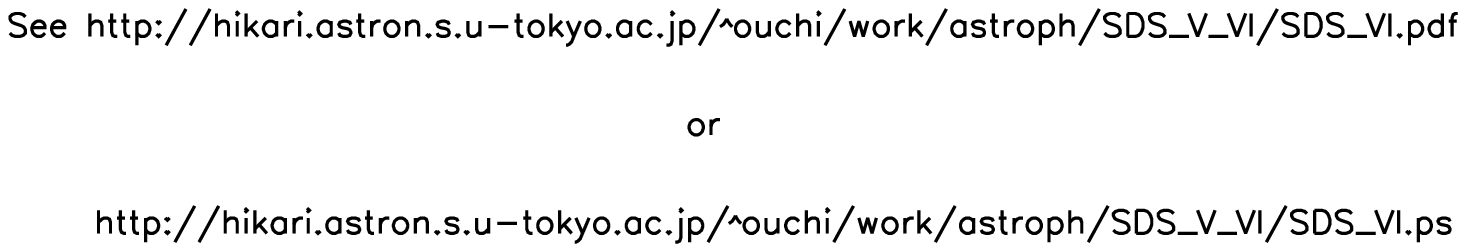}
\caption{
Redshift distributions, $N(z)$, of our
LBG samples obtained in SDS V.
The solid and dashed lines indicate
the number-weighted completeness of the LBG samples
in the SDF and the SXDF, respectively.
Since the tighter selection criteria
(eq. \ref{eq:lbgselection_SXDFVizLBG}) are applied to 
$Viz$-LBGs of the SXDF, 
their selection window is narrower than 
that of the $Viz$-LBGs in the SDF.
\label{fig:completeness_apjpaper}}
\end{figure}

We estimate the redshift distribution, completeness, 
and contamination of the LBG samples using
the probability maps obtained from the simulations.
The redshift ranges are found to be $z=4.0\pm0.5$ for $BRi$-LBGs, 
$z=4.7\pm0.5$ for $Viz$-LBGs, and $z=4.9\pm0.3$ for $Riz$-LBG.
Figure \ref{fig:completeness_apjpaper} shows the number-weighted 
redshift distributions.
The ratio of contaminants to sample objects in number is 1\%
for $BRi$-LBGs, 26\% for $Viz$-LBGs, and 40\% for $Riz$-LBGs in the SDF.
The completeness of each sample is 40-60\% in the most efficient
redshift range. The small ratio of contaminants in $BRi$-LBGs is 
due to the fact that the $BRi$-LBG sample includes many
faint LBGs beyond $M^*$, which are much more numerous than
foreground contaminants.

\subsection{Spectroscopic Follow-up Observations}

In order to examine how well these 
selection criteria isolate true LBGs,
we use 85 spectroscopically identified objects at $0<z<5$
in our fields obtained by 
\citet[hereafter SDS III]{kashikawa2003},
SDS IV, and SDS V.
We find that 6 out of the 85 objects are LBGs at $z>3.5$,
and that no spectroscopically identified 
low-$z$ interloper is included in our LBG
samples. Thus, our LBG samples are thought to be reliable.

\section{Spatial Distribution}
\label{sec:spatial}

\subsection{Sky Distribution}
\label{sec:spatial_sky}

Figures \ref{fig:sdf_dist_z4}-\ref{fig:sxdf_dist_z5} show 
the sky distributions of $BRi$-LBGs, $Viz$-LBGs, and $Riz$-LBGs
detected in the SDF and the SXDF. 
We plot all $\langle z\rangle \simeq 5$
objects, i.e., $Viz$-LBGs and $Riz$-LBGs,
in a single frame; Figure \ref{fig:sdf_dist_z5} is
for the SDF and Figure \ref{fig:sxdf_dist_z5} is for the SXDF.
In addition to these LBGs, we also plot the positions of
87 Lyman $\alpha$ emitters (LAEs) at $z=4.86\pm 0.03$ found by
narrow-band observations (SDS II) in Figure \ref{fig:sdf_dist_z5}.
As found in Figures \ref{fig:sdf_dist_z5} and \ref{fig:sxdf_dist_z5},
some of the objects are selected by
both $Viz$- and $Riz$-LBG (and/or LAE at $z=4.86$) selection criteria.
This is because these criteria select objects in similar
redshift ranges: $Viz$-LBGs at
$z=4.7\pm0.5$, $Riz$-LBGs at $z=4.9\pm0.3$, and
LAEs at $z=4.86\pm0.03$.
Table \ref{tab:detection_coincide} shows
the numbers of objects identified by
two selection criteria. These numbers apparently
look quite small.
We examine whether these numbers are reasonable.
The number of objects selected by two criteria is predicted by:
\begin{equation}
N_{\rm EST}\simeq (N[{\rm SEL1}]-N_{c}[{\rm SEL1}])\times
\frac{\Delta z[{\rm SEL2}]}{\Delta z[{\rm SEL1}]}\times \bar{C} [{\rm SEL2}],
\label{eq:distribution_coincide}
\end{equation}
where $N[{\rm SEL1}]$ is the total number of objects selected by
the first criterion, $N_{c}[{\rm SEL1}]$ 
%
%
is the total number of contaminants in the objects selected by
the first criterion, $\Delta z[{\rm SEL2}]$ and 
$\Delta z[{\rm SEL1}]$ are the approximate 
redshift ranges of the second and the first criteria,
and $\bar{C}[{\rm SEL2}]$ is the number-weighted mean completeness
of the second criterion.
For this calculation, we use the contamination, completeness, 
and redshift distribution derived in 
SDS V for LBGs and those derived in SDS II for LAEs.
See SDSV for details.
The estimated numbers of objects 
identified by two selection criteria are
given in the fourth column of Table \ref{tab:detection_coincide}, 
and are consistent with the actual observed numbers
within Poisson errors.
This coincidence not only explains
the relatively small numbers of objects identified
by two selection criteria, but also ensures the accuracy of
the contamination, completeness,
and redshift distribution obtained
by SDS V and SDS II.

\subsection{Tests for Detection Inhomogeneity}
\label{sec:spatial_tests}

\begin{figure}
\plotone{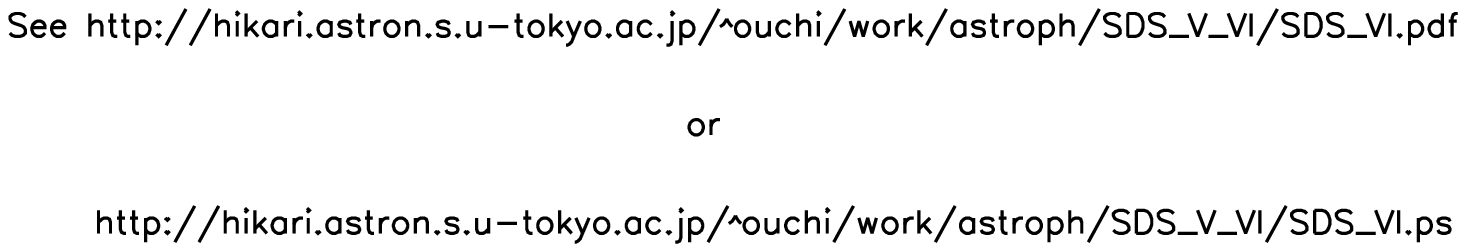}
\caption{
Sky distribution of $z\sim 4$ $BRi$-LBGs detected in the SDF.
The filled circles indicate $BRi$-LBGs; larger symbols
mean brighter magnitudes ($2''\phi$ 
aperture magnitudes in $i'$ band), 
as defined in the figure.
The thick solid line outlines the area of $B$, $V$, $i'$,
and $z'$ images after regions of low S/N are trimmed.
The blank area in the lower-right corner outlined by the
dotted line does not have $R$-band data and thus the 
$BRi$-LBG selection is not applied.
The shaded regions are large masked areas ($>2000$ arcsec$^2$)
where we do not attempt the object detection because of the
presence of bright stars.
%
%
%
The small panel inserted in the lower right corner presents 
a reduction of the main panel on the scale of one to five.
The ten boxes drawn by dotted lines indicate the positions of 
ten CCDs of the Suprime-Cam.
The real size of individual CCDs is $6.'7 \times 13.'3$.
%
%
%
The projected scale of $10 h^{-1}_{100}$ Mpc at $z=4.0$ is shown
in the bottom left of the panel.
North is up and east is to the left in this image.
\label{fig:sdf_dist_z4}}
\end{figure}

\begin{figure}
\plotone{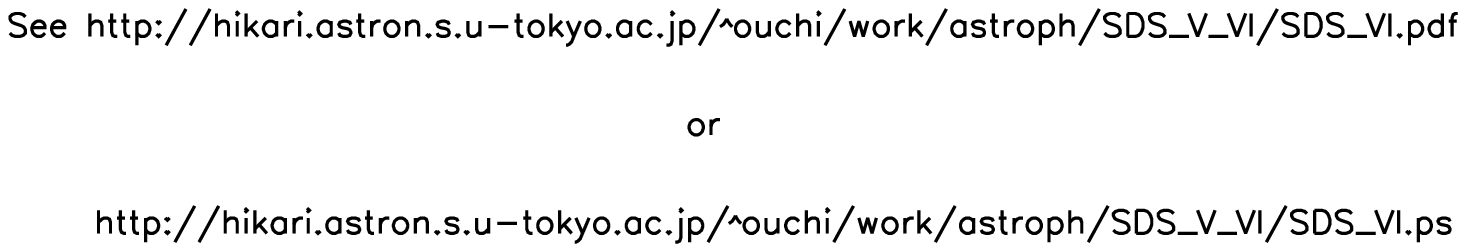}
\caption{
Sky distributions of $z\sim 5$ $Viz$- and $Riz$-LBGs detected in the SDF.
The black filled (red open) circles present
$Viz$-LBGs ($Riz$-LBGs), and the size of the circles 
increases with the apparent brightness in the $z'$ band,
as defined in the figure.
The positions of Lyman $\alpha$ emitters
(LAEs) at $z=4.86\pm0.03$ obtained by SDS II 
are also plotted by blue crosses.
The thick solid line is the whole field of view (FoV) of
our observations in the SDF, and the dashed line
indicates the border of the area of the $R$-band image.
The shaded regions are large masked areas ($>2000$ arcsec$^2$)
where we do not attempt the object detection because of the
presence of bright stars.
%
%
%
%
The projected scale of $10 h^{-1}_{100}$ Mpc at $z=4.7$ is shown
in the bottom left of the panel.
North is up and east is to the left in this image.
\label{fig:sdf_dist_z5}}
\end{figure}

\begin{figure}
\plotone{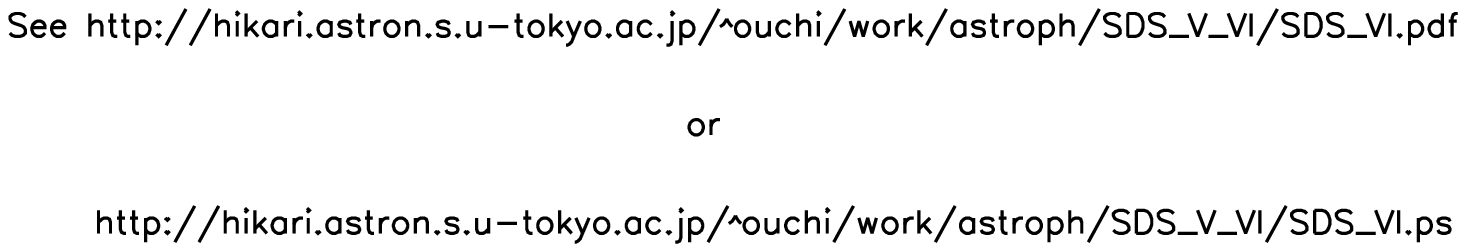}
\caption{
Same as Figure \ref{fig:sdf_dist_z4}, but for $BRi$-LBGs
detected in the SXDF.
North is up and east is to the left in this image.
%
%
%
The small panel inserted is a reduction of the main panel 
on the scale of one to five.
The positions of nine CCDs are shown by dotted lines.
%
%
%
\label{fig:sxdf_dist_z4}}
\end{figure}

\begin{figure}
\plotone{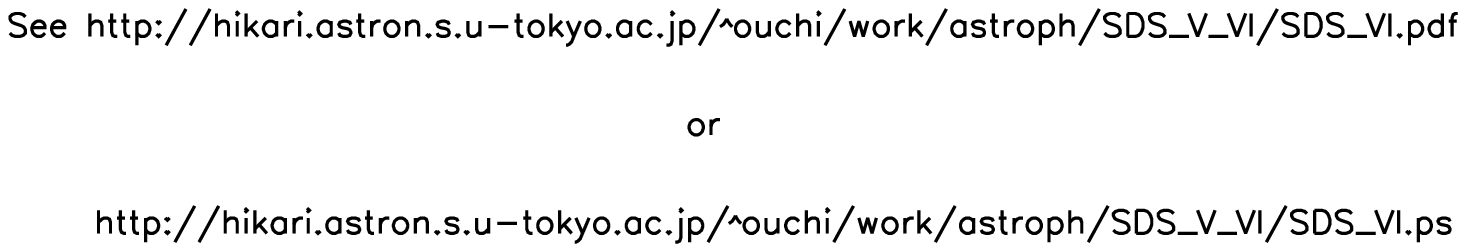}
\caption{
Same as Figure \ref{fig:sxdf_dist_z4}, but for 
$Viz$- and $Riz$-LBGs detected in the SXDF.
Note that a search for LAEs has not been
carried out in the SXDF.
%
%
%
%
North is up and east is to the left in this image.
\label{fig:sxdf_dist_z5}}
\end{figure}

Figures \ref{fig:sdf_dist_z4}-\ref{fig:sxdf_dist_z5} show
somewhat inhomogeneous distributions of LBGs.
The surface density of LAEs has a large gradient over the
whole image.
Prior to investigating their clustering properties,
we examine whether or not 
these inhomogeneities come from spatial inhomogeneities
of the detection efficiency and/or photometric accuracy 
in the images.
%
%
We find that the photometric zero points 
are accurate within 0.1 mag over the whole field of view
for any bandpass, since PSF-like objects are found to
make a single sharp stellar locus 
in any two-color plane as shown in SDS V. 
%
%

%
%
Then we examine whether there are spatial differences in the
source detection efficiency
by (1) calculating the number densities of all detected
objects in small ($10'\times 10'$) areas covering the
survey region,
(2) measuring the limiting magnitudes
in 2700 small ($40''\times40''$) areas for each of
the $B$, $V$, $R$, $i'$, $z'$, and $NB711$ images, 
and (3) estimating the detection
completeness of LBGs and LAEs from Monte Carlo simulations
in the same manner as used in SDS II.
We find no clear sign of inhomogeneity in either of (1)-(3)
for the SDF data. On the other hand, we find a systematic difference
in the detection limits for the SXDF data with a 
0.2 magnitude level on arcminute scales.
This is probably because the majority of the SXDF data were taken 
in 2000, when the CCDs installed were a mixture of products of 
two companies and so had a large variety in quantum efficiency.
%
%
%
%
%
Figures \ref{fig:sdf_dist_z4}-\ref{fig:sxdf_dist_z5} show, however,
that there is no correlation between the source distribution
and the CCD positions.
%
%
%
This demonstrates that the inhomogeneities seen in
Figures \ref{fig:sdf_dist_z4}-\ref{fig:sxdf_dist_z5}
are real, although distributions of LBGs in the SXDF 
on arcminute scales could be contaminated 
by the somewhat large inhomogeneities 
in the limiting magnitude.
%
%

\subsection{Angular Correlation Function}
\label{sec:spatial_angular}

In order to quantitatively 
measure the inhomogeneity of spatial distribution,
we derive the angular two-point correlation function
$\omega(\theta)$.
According to \citet{landy1993},
the angular two-point correlation function is calculated by:

\begin{equation}
\omega_{obs}(\theta)
  = [DD(\theta)-2DR(\theta)+RR(\theta)]/RR(\theta),
\label{eq:landyszalay}
\end{equation}
where $DD(\theta)$, $DR(\theta)$, and $RR(\theta)$ are numbers of
galaxy-galaxy, galaxy-random, and random-random pairs normalized by
the total number of pairs in each of the three samples.
We create the random sample composed of 100,000 sources with the same
geometrical constraints as of the data sample. 
The formal error 
\footnote{The formal error does not 
include a sample variance, which is
caused from field-to-field variations.
}
in $\omega$($\theta$) is described by

\begin{equation}
\sigma_\omega = \sqrt{[1+\omega_{obs}(\theta)]/DD} 
\label{eq:acorr_error}
\end{equation}
%
%
%
\citep{hewett1982}. Its $2\sigma$ error is comparable
to the error obtained with a bootstrap resampling of
the data \citep{baugh1996}.
%
%
%
The real correlation function $\omega(\theta)$ is
offset from the observed function
by the integral constant ($IC$: Groth \& Peebles 1977) as,
\begin{equation}
\omega(\theta)= \omega_{obs}(\theta)+IC.
\label{eq:integralconstant}
\end{equation}
We apply the correction for the IC.
\footnote{
If we assume $\beta=0.8$,
$IC/A_\omega^{\rm raw}$ is estimated to be 
$\simeq 0.006$ for our LBG samples,
where $A_\omega^{\rm raw}$ and $\beta$ 
are the correlation amplitude and the power-law index
shown in eq. (\ref{eq:a_w}).
}

\begin{figure}
\plotone{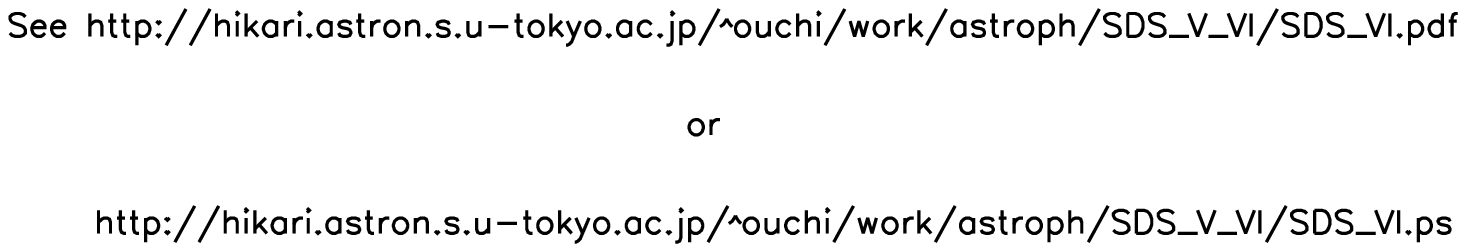}
\caption{
Angular correlation functions, $\omega(\theta)$, of 
$BRi$-LBGs (top), $Viz$-LBGs (middle), and $Riz$-LBGs (bottom). 
The filled circles are for the SDF, 
and the open circles in the top panel are for the SXDF.
No significant signal is detected for either $Viz$-LBGs
or $Riz$-LBGs in the SXDF.
The solid lines show the best-fit power law function,
$\omega(\theta)=A_\omega^{\rm raw} \theta^{-\beta}$, where
$\beta$ is fixed at 0.8. Solid line of the top panel
is the best-fit function both for the SDF and the SXDF data.
\label{fig:acorr_comb}}
\end{figure}

Figure \ref{fig:acorr_comb} shows the angular correlation
functions, $\omega(\theta)$, of $BRi$-LBGs (top panel),
$Viz$-LBGs (middle panel), and $Riz$-LBGs (bottom panel). 
%
%
We measure the angular correlation function 
less than $\simeq 15$ arcmin scales, 
since uncertainties in the measurements increase 
largely over $\sim 15$ arcmin scales, which are comparable 
to a half of the image size.
%
%
In Figure \ref{fig:acorr_comb}, the filled
circles are for the SDF and the open circles in the top panel are for 
the SXDF. 
%
%
We find significant clustering signals for $BRi$-LBGs 
in both the SDF and the SXDF.
%
%
%
The data points of the SDF and SXDF samples  
agree within the $2\sigma$ level except for
the points at $\theta \simeq 30''$ and $\simeq 140''$, 
where the measurements of the SXDF are significantly lower than
those of the SDF.
The reason for these discrepancies is not clear. 
However, it could be related to the worse
quality of the SXDF data than of the SDF data.
In any case, the SDF sample is deeper than the SXDF sample 
and does not suffer from any detectable 
(artificial) inhomogeneity in the data.
Hence, we believe that the angular correlation measured for 
the SDF sample is more reliable.
%
%
%
We also find clear clustering signals for $Viz$-LBGs in the SDF.
There is a marginal signal for $Riz$-LBGs in the SDF, and
no signal is detected for $Viz$-LBGs and $Riz$-LBGs in the SXDF.
This is mainly because the numbers of objects in these samples 
are considerably smaller than those in the other samples.
%
%

We fit a single power law,
\begin{equation}
\omega(\theta)=A_\omega^{\rm raw} \theta^{-\beta},
\label{eq:a_w}
\end{equation}
to the data points. 
%
%
For $BRi$-LBGs, a fit is made to a combination of the data points 
of the SDF and the SXDF 
(A fit to the SDF data alone gives almost the same results 
since the errors in the data points of the SDF sample are 
much smaller).
%
%
Errors in $A_\omega^{\rm raw}$ and $\beta$ are found to be very large
%
%
except for $BRi$-LBGs, $\beta=0.90^{+0.11}_{-0.07}$ 
and $A_\omega^{\rm raw}=2.3^{+0.9}_{-0.6}$ 
%
%
(Table \ref{tab:acorr_comb}, line 1).
This is probably due to the large sample size of
$BRi$-LBGs.
%
%
The value for $\beta$ for $BRi$-LBGs is close to 
the {\lq}fiducial{\rq} value, 0.8. 
Thus, we use $\beta \equiv 0.8$ in the rest of this paper.
Adopting $\beta=0.91$ changes the results very little.
%
%
%
Lines 2-14 of Table \ref{tab:acorr_comb} present
the best fit values of $A_\omega^{\rm raw}$
with $\beta=0.8$. 
The correlation amplitudes $A_\omega^{\rm raw}$ of 
$BRi$-LBGs and $Viz$-LBGs
are as high as at the
$\sim10 \sigma$ and $\sim2.5 \sigma$ levels, respectively.
Hence, their clustering signals are significant.
On the other hand,
the $A_\omega^{\rm raw}$ value of
$Riz$-LBGs is only at the
$< 1 \sigma$ significance level.
%
%
Thus, in the following discussion,
we adopt the $A_\omega^{\rm raw}$ value of SDF $Viz$-LBGs 
for the clustering amplitude of $z=5$ LBGs.
%
%
%

Foreground contamination to a galaxy sample 
dilutes the apparent clustering amplitude of galaxies. 
When the fraction of contaminants is $f_c$,
the apparent $A_\omega^{\rm raw}$ value  
can be reduced by a factor of up to $(1 - f_c)^2$.
The true correlation amplitude, $A_\omega$, is 
given by
\begin{equation}
A_\omega = \frac{A_\omega^{\rm raw}}{(1-f_c)^2}.
\label{eq:clustering_dilution}
\end{equation}
This is the maximum reduction of the correlation
amplitude that occurs when the contaminants 
are not at all clustered.
In reality, the contaminants in our sample will be the sum of 
foreground galaxies at various redshifts, and thus would be 
clustered very weakly, if any, on the sky.
Thus we use eq. (\ref{eq:clustering_dilution}) to compute
$A_\omega$ for our LBG samples.
We use $f_c$ obtained by simulations described in 
SDS V for LBGs.
Table \ref{tab:acorr_comb} also gives $A_\omega$ values.

The effect of field-to-field variations in our samples is 
probably modest for LBGs, since 
we find that angular correlation functions of
$BRi$-LBGs in the SDF and the SXDF agree moderately well
within error bars (Figure \ref{fig:acorr_comb}).
This is because our LBG samples probe large comoving volumes;
$1.6\times 10^6 h_{70}^{-3}$Mpc$^{3}$ for $BRi$-LBGs and
$1.7\times 10^6 h_{70}^{-3}$Mpc$^{3}$ for $Viz$-LBGs.
On the other hand, the surveyed volume of LAEs at $z=4.9$
shown in SDS II is only $9.0\times 10^4 h_{70}^{-3}$Mpc$^{3}$.
We may have to consider the cosmic variance for the LAE sample.

%
%
%
Since we have a large number of
LBGs at $z=4$ in the SDF, we make subsamples which are 
divided by magnitude or color.
We use two observational quantities, $i'$ magnitude
and $i'-z'$ color, and make three types of subsamples;
a subsample composed of objects 
with $M\lesssim M^*$, subsamples divided by $M$, 
and subsamples divided by $E(B-V)$. 
Then we calculate the angular correlation functions of
these subsamples in the same manner as for the whole LBG samples
shown above. 
We describe the details of the analysis 
in subsections \ref{sec:spatial_evolution} 
and \ref{sec:spatial_segregation}, 
and summarize the selection criteria and the results in Table 3.
%
%
%
%
%
%
We do not apply this analysis to the SXDF $BRi$-LBG sample, 
since its limiting magnitude is brighter than the SDF sample's 
(and so the size of the SXDF sample is about half 
that of the SDF sample) 
and the SXDF sample could 
suffer from the inhomogeneity of the limiting magnitude
over the image. 
%
%

\subsection{Correlation Length}
\label{sec:spatial_correlation}

Since the angular correlation function
shows clustering properties of galaxies projected on
the sky, $\omega(\theta)$ reflects a combination
of the redshift distribution of the selected galaxies and 
the intrinsic clustering in the three dimensional space, i.e.,
the spatial correlation function, $\xi_g$.
The spatial correlation function of galaxies is 
usually expressed by a power law as
\begin{equation}
\xi_g=(r/r_0)^{-\gamma},
\label{eq:3dcorrelation}
\end{equation}
where $r$ is the spatial separation between two objects,
$r_0$ is the correlation length, and $\gamma$ is the slope
of the power law.
The correlation length, $r_0$, 
is related to the correlation amplitude, $A_\omega$, 
with the integral equation
called Limber equation \citep{peebles1980,efstathiou1991},
\begin{equation}
      A_{\omega}= C r_0^\gamma
      \int_{0}^{\infty} F(z) D_\theta^{1-\gamma}(z)N(z)^2g(z) dz 
      \left[ \int_{0}^{\infty} N(z) dz \right]^{-2},
 \label{eq:limber}
\end{equation}
where $F(z)$ 
\footnote{
Assuming that the clustering pattern 
is fixed in comoving coordinates in the
redshift range of our sample, 
we take the functional form,
$F(z)={(1+z)/(1+z_c)}^{-(3+\epsilon)}$,
where $z_c$ is the central redshift of the sample
and $\epsilon=-1.2$. 
The effect of the change in $\epsilon$ over
$0<\epsilon<-3$ on $r_0$ is, however, very small.
}
describes the redshift dependence of $\xi(r)$, 
$D_\theta(z)$ is the angular diameter distance, 
$N(z)$ is the redshift distribution of objects,
$
g(z)=H_0/c[(1+z)^2(1+\Omega_m z 
    + \Omega_\Lambda((1+z)^{-2}-1))^{1/2}],
$
and $C$ is a numerical constant, 
$C=\sqrt{\pi} \Gamma[(\gamma -1)/2]/\Gamma(\gamma/2)$.
The slope $\beta$ of the angular correlation function
is related to $\gamma$ by 
\begin{equation}
\gamma=\beta+1.
\label{eq:gamma}
\end{equation}
We need the redshift distributions of our LBGs to
derive $r_0$.
We adopt the ones obtained
in section \ref{sec:data}.

The correlation lengths thus obtained are summarized in
Table \ref{tab:r0_comb}. In Table \ref{tab:r0_comb},
we show both $r_0$ and $r_0^{\rm raw}$, which
are the correlation lengths obtained from
$A_\omega$ and $A_\omega^{\rm raw}$. In the following
discussion, we regard $r_0$ as the best estimate value.
We find $r_0=
4.1^{+0.2}_{-0.2}$ and $5.9^{+1.3}_{-1.7} h_{100}^{-1}$Mpc for
all LBGs at $z=4$ and $5$, respectively. 
\citet{ouchi2001a} calculate the correlation length of 
LBGs at $z=4$ in the same manner
as above, but using a sample constructed based on different selection
criteria. They report that LBGs at $z=4$ with $i'<26$ have
$r_0 = 3.3^{+0.6}_{-0.7} h_{100}^{-1}$Mpc 
($r_0^{\rm raw}=2.7^{+0.5}_{-0.6} h_{100}^{-1}$Mpc).
This value is consistent with
that obtained from our LBG sample.
The correlation lengths we obtain ($r_0=3-8 h_{100}^{-1}$ Mpc) 
correspond to projected
angular scales of $100''-300''$ at $z=4-5$.
These scales fall in
the range where the power-law fitting is excellent 
(Figure \ref{fig:acorr_comb}). 

\subsection{Evolution of Correlation Length and Galaxy Distribution Bias}
\label{sec:spatial_evolution}

We discuss the redshift evolution of
the correlation length.
%
%
Since the clustering amplitude
(i.e., correlation length) depends on the brightness of galaxies 
(see section \ref{sec:spatial_segregation_lbg_uv}),
we compare the correlation lengths of LBGs whose luminosities
are $L\gtrsim L^*$. We calculate the correlation length of LBGs
with $L\gtrsim L^*$ for $z=4$ and $5$. 
We make a subsample composed of 357 
LBGs at $z=4$ with $i'<25.3$ corresponding to $M<-20.8$ ($\simeq M^*$),
and calculate the correlation length to be
$r_0= 5.1^{+1.0}_{-1.1} h_{100}^{-1}$ Mpc
(line 2 of Table \ref{tab:r0_lstar}).
Since the limiting magnitude of $z=5$ LBGs
is $z'=25.8$ which corresponds to $M=-20.5$ ($\simeq M^*$),
we regard the results of all $z=5$ LBGs 
(line 2 of Table \ref{tab:r0_comb})
as those of $L\gtrsim L^*$ LBGs
(line 3 of Table \ref{tab:r0_lstar}).
Thus the correlation length of $L\gtrsim L^*$ LBGs at $z=5$ 
is $r_0=5.9^{+1.3}_{-1.7} h_{100}^{-1}$ Mpc.
We also show in Table \ref{tab:r0_lstar} 
the correlation lengths of 
LBGs with $L\gtrsim L^*$ at $z=3$
given by \citet{giavalisco2001} and
of all LAEs at $z=4.9$
obtained by SDS II.
%
%

The top panel of Figure \ref{fig:z_r0_bias} plots 
$r_0$ as a function of redshift, together
with $r_0$ of the underlying dark matter calculated by
the non-linear model of \citet{peacock1996}.
The correlation lengths of LBGs
with $L\gtrsim L^*$ 
is almost constant around $r_0 \sim 5 h_{100}^{-1}$Mpc
at $z=3-5$, although it might
increase slightly with redshift. It is also found that
the correlation amplitudes of LBGs (and LAEs) are much
higher than that of the underlying dark matter. 
We define the galaxy-dark matter clustering bias
of the large scale ($=8 h_{100}^{-1} {\rm Mpc}$) as 
\begin{equation}
b_g=\sqrt{\frac{\xi_g(r=8 h_{100}^{-1} {\rm Mpc})}{\xi_{\rm DM}(r=8 h_{100}^{-1} {\rm Mpc})}},
\label{eq:bias}
\end{equation}
where $\xi_{DM}$ is the two-point correlation function of the underlying
dark matter. 

We also calculate $r_0$ of the dark matter using linear theory (e.g.,
\citealt{peacock1994}), which is plotted 
in Figure \ref{fig:z_r0_bias} by a dashed line, 
and find a negligible difference from 
the calculation based on the non-linear model.
This is because
the effect of non-linearity is very small on scales of $r\geq r_0$.
\citet{jenkins1998} show that 
linear theory traces the results of $N$-body simulations
above $r_0$
at $z=0$ where the non-linear effect is larger
than at high redshifts.

Using linear theory,
we calculate $b_g$ by eq. (\ref{eq:bias})
from $r_0$ (eq. \ref{eq:3dcorrelation}), 
$\gamma$ (eq. \ref{eq:gamma}), and
$\xi_{\rm DM}$.
We explicitly present the relation between $r_0$ and $b_g$
given in eq. (\ref{eq:bias}):
\begin{equation}
b_g=\sqrt{\frac{[(8h_{100}^{-1}{\rm Mpc})/r_0]^{-\gamma}}
{\xi_{\rm DM}(r=8h_{100}^{-1}{\rm Mpc})}}.
\label{eq:bias_detail}
\end{equation}
We show values of $b_g$ in Table \ref{tab:r0_lstar} 
for LBGs with $L\gtrsim L^*$ and all LAEs
(and Table \ref{tab:r0_comb} for all samples), and
the bottom panel of Figure \ref{fig:z_r0_bias}
plots $b_g$ against redshift for LBGs with $L\gtrsim L^*$
and all LAEs. These LBGs and LAEs are biased against
dark matter by $b_g\sim 3-5$, and the bias becomes stronger 
at higher redshifts.
This increase in LBGs' bias can be regarded as a piece of evidence 
supporting the biased-galaxy formation scenario (e.g., \citealt{baugh1999}).

\subsection{Segregation of Clustering Amplitudes}
\label{sec:spatial_segregation}

\subsubsection{Clustering Segregation with UV-Continuum Luminosity}
\label{sec:spatial_segregation_lbg_uv}

Since we have a large number 
of LBGs at $z=4$,
we examine the luminosity dependence of clustering amplitude
($r_0$ and $b_g$) using four subsamples binned according to the source 
brightness ($\Delta m = 1.5$).
%
%
%
We calculate the angular correlation functions for 
the four subsamples,
and estimate $r_0$ and $b_g$ in the same manner as described 
in sections \ref{sec:spatial_correlation} and \ref{sec:spatial_evolution}, 
but applying to each subsample its own redshift distribution 
which is determined 
by simulations described in SDS V (Figure 12 of SDS V).
%
%
%
%
Lines 6-9 of Table \ref{tab:acorr_comb} and 
lines 4-7 of Table \ref{tab:r0_comb}
give the best-fit parameters
of angular correlation function together with
$r_0$ and $b_g$ for these four subsamples. 
Figure \ref{fig:bias_norberg} plots $r_0$ and $b_g$
for the four subsamples as a function of magnitude.
It is found from Figure \ref{fig:bias_norberg} 
that brighter LBGs are clustered more
strongly, which is a tendency similar to 
that found for present-day galaxies \citep{norberg2002b} and 
LBGs at $z=3$ \citep{giavalisco2001}.
It is also found that this luminosity dependence of $r_0$ is
weaker for fainter galaxies.
We fit the data points by a linear function of luminosity ($L$):
\begin{equation}
b_g/b_g^* = a + (1-a) L/L^*,
\label{eq:norbergslaw}
\end{equation}
where $a$ is a free parameter,
$b_g$ is the galaxy-dark matter bias,
and $b_g^*$, 
which is also a free parameter,
is the bias of galaxies with
luminosity, $L^*$ (Norberg's law; \citealt{norberg2002b}).
We obtain $b^*=2.8$ and $a=0.58$
with $L^*$ being fixed at $M_{\rm 1700}^*=-21.0$ (or $i^*= 25.1$),
which is obtained in SDS V.
The value of $a$ for $z=4$ LBGs is smaller than that for
present-day galaxies, $a=0.85$ \citep{norberg2002b}.
It may indicate that the clustering amplitude of $z=4$ LBGs
depends more strongly on luminosity, and/or that 
the luminosity dependence of clustering is different 
between rest-frame UV luminosity (for LBGs) and optical luminosity
(for present-day galaxies).

\subsubsection{Clustering Segregation with Dust Extinction}
\label{sec:spatial_segregation_lbg_dust}

Strong clustering has been reported for SCUBA sources 
($r_0=12.8\pm4.5\pm3.0h_{100}^{-1}$Mpc; \citealt{webb2003b}).
Since SCUBA sources are thought to be dust rich starburst galaxies,
the clustering amplitude of LBGs as a function of dust extinction
may give a hint for the relation between SCUBA sources
and LBGs.
We calculate the clustering amplitude for subsamples 
of LBGs at $z=4$ binned with $E(B-V)$.
We estimate $E(B-V)$ of LBGs from their colors 
with equations shown in SDS V.
We make these subsamples from 357 bright LBGs 
with $i'<25.3$ ($\simeq M^*$)
so that we accurately measure colors, $i'-z'$, and thus
obtain $E(B-V)$. We make five subsamples whose
$E(B-V)$ values range from $-0.2$ to $0.6$ corresponding
to $i'-z'$ color from $-0.18$ to $0.50$.
\footnote{
The relation between $E(B-V)$ and $i'-z'$ is
given by $E(B-V)\simeq 0.0162+1.18(i'-z')$
(see SDS V).
}
Lines 10-14 of Table \ref{tab:acorr_comb} and 
lines 8-12 of Table \ref{tab:r0_comb}
present the best-fit parameters
of angular correlation functions, together with
$r_0$ and $b_g$, for
these five subsamples. 
We find that 
the most dusty subsample ($0.2<E(B-V)<0.6$, i.e., $E(B-V)\sim 0.4$)
has the strongest clustering: 
$r_0=8.2^{+2.9}_{-4.2} h_{100}^{-1}$ Mpc.
Although the errors are large, it may suggest that
very dusty LBGs have a connection to SCUBA sources.
%
%
%

On the other hand, \citet{ouchi1999} find that
no LBG identified in the HDF-N is detected 
by deep SCUBA observations \citep{hughes1998}.
There are three possible explanations for this finding.
First, dusty LBGs bright enough to be detected with SCUBA 
may be so rare that the probability of finding such an LBG
in a small area like the HDF-N is very low.
Second, the redshift distributions of LBGs and SCUBA sources
may overlap little with each other.
Indeed, SCUBA sources are generally located at $z=1.9-2.8$
\citep{chapman2003}, while LBGs are found at redshifts 
higher than $z\simeq 2.5$.
Third, LBGs may have a closer relation with EROs 
than with SCUBA sources.
EROs are thought to be a mixture of old galaxies and 
dusty star-forming galaxies.
Although we assume that the red $i'-z'$ color of LBGs is 
due to dust extinction, the reddest LBG subsample might 
include a large number of old galaxies 
which have optical-to-infrared colors close to those of EROs. 
In reality, \citet{mccarthy2001} find that 
the correlation length of EROs at $z\sim 1$ 
is $9-10 h_{100}^{-1}$ Mpc, 
comparable to that of the reddest LBG subsample. 
%
%
%

\section{Galaxy and Structure Formation based on the Hierarchical Clustering Picture}
\label{sec:galaxy}

\subsection{Joint Analysis of Luminosity Functions and Clustering Amplitudes
based on the Cold Dark Matter Model}
\label{sec:galaxy_joint}

 We estimate the mass of dark halos which host LBGs and LAEs using
the analytic models given by \citet{sheth1999} and \citet{sheth2001}.
They are derived from a fit to
the results of GIF simulations (GIF/Virgo collaboration; 
e.g., \citealt{kauffmann1999}).

Prior to estimating the mass of dark halos,
we present an overview of the model of
\citet{sheth1999}.
According to their model (see also \citealt{mo2002}),
the number density of
collapsed dark halos with mass $M$ at redshift $z$ is given by
\begin{equation}
n(M,z)dM = A(1+\frac{1}{\nu'^{2q}})\sqrt{\frac{2}{\pi}}
\frac{\bar{\rho}}{M} \frac{d\nu'}{dM} \exp \left(-\frac{\nu'^2}{2}\right)dM,
\label{eq:mass_func_shethtormen}
\end{equation}
where $\nu'=\sqrt{a}\nu$, $a=0.707$, $A\simeq 0.322$, and $q=0.3$.
Here, $\nu$ is defined by
\begin{equation}
\nu\equiv \frac{\delta_c}{D(z)\sigma(M)},
\label{eq:nu}
\end{equation}
where 
$D(z)$ is the growth factor, $\sigma (M)$ is the rms of
the density fluctuations on mass scale $M$, and
$\delta_c=1.69$ represents the critical amplitude
of the perturbation for collapse. 
We calculate the growth factor following \citet{carroll1992}.
$\sigma (M)$ is calculated from the power spectrum
whose power law index is $n=1$,
using the transfer function
given by \citet{bardeen1986}.
The number density of collapsed dark halos at a given
redshift, i.e., the mass function of dark halos,
is calculated using 
eqs. (\ref{eq:mass_func_shethtormen}) and (\ref{eq:nu}).
The resulting mass function at $z=4$ is shown 
in Figure \ref{fig:mass_func_diff_z4}.

In the model of \citet{sheth1999}, the bias of dark halos
against the dark matter, $b_{\rm DH}$, is calculated by
\begin{equation}
b_{\rm DH} = 1+\frac{1}{\delta_c} \left[\nu'^2+b\nu'^{2(1-c)}-
\frac{\nu'^{2c}/\sqrt{a}}{\nu'^{2c}+b(1-c)(1-c/2)} \right],
\label{eq:bias_sheth}
\end{equation}
where $b=0.5$, $c=0.6$, and the other parameters
($\nu'$, $\delta_c$, and $a$) are the same as
those in eqs. (\ref{eq:mass_func_shethtormen}) 
and (\ref{eq:nu}) \citep{sheth2001}.
Since $\nu'$ is a function of redshift and mass,
the mass of a dark halo is estimated 
from eqs. (\ref{eq:nu})and (\ref{eq:bias_sheth}),
once the bias of the dark halo
and the redshift are given.
Since the bias values of LBGs, $b_g$,
are measured at a large separation of $8 h_{100}^{-1}$ Mpc,
these values reflect the bias of dark halos (which host LBGs)
rather than the bias of the 
galaxy distribution in dark halos, i.e., $b_g\simeq b_{\rm DH}$.
Assuming that galaxies with a given brightness
are hosted 
by dark halos of a specific mass (i.e., there is a one-to-one
correspondence between the galaxy luminosity and the halo mass),
we estimate the average mass of dark halos, $\langle M \rangle$, 
for galaxies of a given luminosity range
from the bias values obtained in section \ref{sec:spatial_evolution}.
We also calculate the number density of galaxies
from the luminosity functions (LFs)
given in SDS V. Note that, since the LFs are derived
by correcting completeness and contamination for the 
number counts of LBGs, these number densities do not
suffer from these systematic observational errors.
The ticks in Figure \ref{fig:bias_norberg} show 
the halo masses and the number densities for $z=4$ LBGs
thus obtained. 
We then plot in Figure \ref{fig:mass_func_diff_z4} 
the number density of
LBGs against the mass of hosting dark halos for each luminosity bin,
together with the mass function of all dark halos existing
at $z=4$ calculated by 
eq. (\ref{eq:mass_func_shethtormen}).

\subsection{Hosting Dark Halos and Halo Occupation Number}
\label{sec:galaxy_hosting}

\subsubsection{Dark Halos of Lyman Break Galaxies at $z=4$}
\label{sec:galaxy_hosting_z4}

Figure \ref{fig:mass_func_diff_z4}
shows that 
dark halos which host LBGs at $z=4$ have
$10^{11}-10^{13} h_{70}^{-1} M_\odot$.
Each filled circle denotes the mass of 
hosting halos for each luminosity bin
(see Table \ref{tab:halo_mass} for the definition of
the luminosity bins); from right to left,
halos hosting LBGs with 
$i'\simeq 24.0$, $24.5$, $25.0$, and $25.5$.
The third circle from right is 
for $i'\simeq 25.0$ ($M_{\rm 1700}\simeq -21$) LBGs,
i.e., for $L^*$-LBGs (see SDS V).
\footnote{
We also have a subsample of $L\gtrsim L^*$ LBGs
which contains all LBGs with $L\gtrsim L^*$
(see section \ref{sec:spatial_angular}).
Note that the samples of $L^*$-LBGs and $L\gtrsim L^*$ LBGs 
are different samples.
}
We find that
bright LBGs ($M\simeq M^*-1$) 
have more massive dark halos 
($4.6^{+4.0}_{-3.4}\times 10^{12} h_{70}^{-1}M_\odot$)
than faint LBGs 
($M\simeq M^*+0.5$; $1.7^{+1.1}_{-0.7}\times 10^{11} h_{70}^{-1}M_\odot$).
A qualitatively similar trend has been found 
for LBGs at $z=3$ \citep{giavalisco2001}.

The most striking feature in Figure \ref{fig:mass_func_diff_z4}
is the discrepancy in number density 
between all dark halos (thin solid line) and
all the data points (filled circles) 
except the brightest one.
We define the occupation number, $N_{\rm occup}$, as the
ratio of the number density of galaxies to that of dark halos.
Table \ref{tab:halo_mass} presents $N_{\rm occup}$ for each luminosity bin
together with the results for galaxies at $z=3$ and $5$,
which are explained in the following sections.
Lines 10-13 of
Table \ref{tab:halo_mass} (and Figure \ref{fig:mass_func_diff_z4})
show that the occupation number
of the brightest LBGs is almost unity, but 
those of fainter LBGs (including $L^*$-LBGs) 
are 3 to 10 times smaller. 
This implies that the majority of low-mass 
($\lesssim 10^{12} h_{70}^{-1} M_\odot$) dark halos have no LBGs.
Since LBGs are star-forming galaxies which are bright in the UV,
it is possible that low-mass dark halos 
have a galaxy which happens to be
between star-forming phases and thus escapes from our
LBG samples.

\subsubsection{Dark Halos of Galaxies at $z=3$}
\label{sec:galaxy_hosting_z3}
In order to investigate the discrepancy in number density
between LBGs and 
all dark halos (Figure \ref{fig:mass_func_diff_z4}),
we carry out the same analyses 
for various galaxy populations at $z\simeq 3$, 
whose number density and
correlation length have been reported in the literature.
We examine $K$-band selected galaxies (FIRES; \citealt{daddi2003}), 
SCUBA sources (SCUBA; \citealt{webb2003b}), and
LBGs \citep{giavalisco2001}.
Here, we note basic 
properties of these three populations. FIRES galaxies
are detected by very deep $K$-band imaging ($K<24$) 
with VLT/ISSAC \citep{daddi2003}. 
Although the sample of FIRES galaxies may
suffer from the field variance
due to its very small survey field ($\sim 4$ arcmin$^2$), 
\citet{daddi2003} claim from their simulations that 
the strong correlation amplitudes they found
are hardly reproduced by the field variance alone.
FIRES galaxies include UV-faint galaxies, i.e., quiescent
in star-formation activity, 
which are not identified by LBG selections \citep{franx2003}.
SCUBA sources are detected by
their strong submillimeter fluxes in 
Canada-UK Deep Submillimeter Survey with JCMT/SCUBA,
and they are thought to be a mixture of
dusty-starburst galaxies and AGNs. 
The sample of LBGs at $z=3$ used in this analysis is taken
from \citet{giavalisco2001} (see also 
\citealt{giavalisco1998,adelberger1998,
arnouts1999,arnouts2002,porciani2002}). 

Figure \ref{fig:mass_func_diff_z3} 
(and lines 1-7 of Table \ref{tab:halo_mass}) 
shows the results for
$z\simeq 3$ galaxies. The mass of dark halos
hosting LBGs ranges over $10^{11}-10^{12} h_{70}^{-1} M_\odot$.
The occupation number of LBGs
is less than unity ($\sim 0.3$: see Table \ref{tab:halo_mass}). 
These features of LBGs at $z=3$ are similar to those at $z=4$.
On the other hand, dark halos of FIRES galaxies have 
$\sim 10^{12} h_{70}^{-1} M_\odot$, which is comparable to 
the mass of LBGs, but the occupation number of FIRES galaxies
is as high as $\simeq 4$. This comparable mass and large occupation number
indicate that dark halos with $\sim 10^{12} h_{70}^{-1} M_\odot$ have
4 galaxies on average, and only about $1/10$ ($\simeq 0.3/4$) 
of them 
have an active-star formation whose star-formation rate (SFR) exceeds
$\sim 5 h_{70}^{-2} M_\odot$yr$^{-1}$ 
(or $\sim 20 h_{70}^{-2} M_\odot$yr$^{-1}$ if 
dust extinction is corrected; see SDS V), 
so that they are identified as LBGs by
their bright UV continuum. 

Halos hosting SCUBA sources are located near the high-mass end of
the dark halo mass function at $z=3$, 
having $\sim3\times 10^{13} h_{70}^{-1} M_\odot$.
The best estimate of the 
occupation number of SCUBA sources is about 4,
though the uncertainty in it is about an order of magnitude
larger than those for other objects
due to large errors in the measurements of
the number density and the correlation length.
The large mass of
dark halos ($\sim 5\times 10^{13} h_{70}^{-1} M_\odot$) 
hosting SCUBA sources indicates that 
SCUBA sources exist only in proto-cluster regions.

\subsubsection{Dark Halos of Lyman Break Galaxies and Lyman $\alpha$ Emitters at $z=5$}
\label{sec:galaxy_hosting_z5}

We then apply our analysis to LBGs at $z=5$ and LAEs at $z=5$
obtained in SDS II,
so as to investigate the evolution of LBGs from $z=3$ to $z=5$
and differences between LBGs and LAEs.
The results are given in lines 14-15 of Table \ref{tab:halo_mass}
and in Figure \ref{fig:mass_func_diff_z5}.
Both LBGs and LAEs at $z=5$ 
are embedded in dark halos with $\sim 10^{12} h_{70}^{-1} M_\odot$.
However, the number densities of LBGs and LAEs are different,
and so
the occupation number of LAEs ($N_{\rm occup}=3.2$)
is about 5 times larger than that of LBGs ($N_{\rm occup}=0.6$).
It indicates that 
the majority of dark halos with $\sim 10^{12} h_{70}^{-1} M_\odot$
do not host LBGs, while hosting multiple LAEs. 
Since the typical brightness of LAEs in SDS II 
is about 2 magnitude fainter in UV continuum
than that of our LBGs at $z=5$, 
galaxies in the LAE sample have about ten times less star-formation 
rates than those in the LBG sample.
This deep detection limit of the LAEs results in 
the larger number density for LAEs.
%
%

Note that the estimated mass of dark halos hosting LAEs may be
smaller than the true values. We use
the best-fit power-law function for 
the angular correlation function 
shown in SDS II
to obtain $r_0$, and calculate the galaxy-dark
matter bias $b_g$ defined at $r=8h_{100}^{-1}$Mpc.
However, the angular correlation function
has a
hump at around $r=8h_{100}^{-1}$Mpc, and
the observed points systematically 
exceed the fitted power law
(see Figure 6 of SDS II).
If the hump of the clustering amplitude at $r\sim 8h_{100}^{-1}$Mpc
is real, the true $b_g$ value is 
larger than that obtained in section 
\ref{sec:spatial_correlation}, resulting in a larger mass
of dark halos ($3.5^{+1.4}_{-0.7}\times 10^{12}$ $h_{70}^{-1} M_\odot$)
and a larger occupation number ($35$).

\subsection{Descendants of Dark Halos Hosting High-$z$ Galaxies}
\label{sec:galaxy_descendants}

Once the mass of dark halos is determined at a given redshift ($z$),
one can calculate analytically the bias of dark halos
for their descendants (and progenitors)
at an arbitrary redshift.
The bias of descendants at the present epoch, $b_{\rm DH}^0$,
is calculated by \citep{sheth2001}:
\begin{equation}
b_{\rm DH}^0 = 1+\frac{D(z)}{\delta_c} \left[\nu'^2+b\nu'^{2(1-c)}-
\frac{\nu'^{2c}/\sqrt{a}}{\nu'^{2c}+b(1-c)(1-c/2)} \right],
\label{eq:bias0_sheth}
\end{equation}
where $D(z)$ is the growth factor,
and the other notations are the same as those in 
eq. (\ref{eq:bias_sheth}).
Assuming that the bias thus obtained represents the
bias of a given dark-halo mass at the present epoch,
we estimate the dark-halo mass of descendants
with eq. (\ref{eq:bias_sheth}).
We first calculate the present-day bias for dark halos
hosting high-$z$ galaxies, 
and then derive the mass of
descendants at $z=0$ from $b_{\rm DH}^0$.

Figure \ref{fig:mass_func_diff_z0} shows the mass of descendants 
of dark halos which hosted high-$z$ galaxies. 
In this figure, the number densities of descendants are
the same as those measured for high-$z$ galaxies, since the analytic
model does not predict the number density of descendants.
In other words, in Figure \ref{fig:mass_func_diff_z0} we assume that
high-$z$ galaxies conserve the number densities until the present
epoch.
It is found from Figure \ref{fig:mass_func_diff_z0} 
that the majority of high-$z$ galaxies are embedded in 
dark halos of $10^{13}-10^{15} h_{70}^{-1} M_\odot$
at the present epoch.
Generally speaking, dark-halo masses of
$10^{13}-10^{15} h_{70}^{-1} M_\odot$ at $z=0$
correspond to the mass scale of clusters and groups. 
Our finding means that the high-$z$ galaxies examined here
become member galaxies of clusters and groups
in the present-day Universe
in a statistical sense. This result is qualitatively
consistent with those found by \citet{nagamine2002} 
who claims from a hydrodynamic simulation that
all bright ($M_V\lesssim -23$) LBGs at $z=3$
evolve into galaxies in clusters or groups at $z=0$, and
that half of LBGs with $-23\lesssim M_V\lesssim -21$ 
fall into clusters or groups. 

In Figure \ref{fig:mass_func_diff_z0},
the dotted line denotes the number density of present-day galaxies,
which is calculated from the occupation-number function
derived by \citet{vandenbosch2003} who used the 2dFGRS data.
Note that this occupation-number function is for
galaxies brighter than $M_{b_g} \sim -19$ (this limiting magnitude is
determined by the depth of the 2dFGRS data).
In other words, the dotted line of 
Figure \ref{fig:mass_func_diff_z0} indicates the number density
of normal bright galaxies.
This number density of present-day normal galaxies
is found to
exceed those of bright LBGs and SCUBA sources. On the other hand,
the number density of 
faint LBGs ($L\lesssim L^*$), FIRES galaxies, and LAEs is 
higher than that of present-day normal galaxies. 
If we assume that all high-$z$ galaxies evolve into
present-day normal galaxies, the excess of
the number density of high-$z$ galaxies 
over that of present-day galaxies 
puts a constraint on the minimum number of mergers
which high-$z$ galaxies have experienced until $z=0$.
We define $N^{\rm min}$(merge) as 
the ratio of the number density
of high-$z$ galaxies to that of present-day
galaxies, and calculate $N^{\rm min}$(merge) for
each of the high-$z$ galaxy populations 
in Table \ref{tab:halo_mass}.
It is found that
$L\lesssim L^*$ LBGs, LAEs, and FIRES galaxies
should experience mergers at least 2-5, 4, and 7 times, respectively,
so as to reduce their number density. 
Part of these high-$z$ galaxies will evolve into massive
cluster galaxies through mergers.
Our findings are consistent with 
results of semi-analytical models. For example,
\citet{governato2001} find that among 12 halos containing
at least one LBG in a proto-cluster region
at $z=3$, 7 halos merge into
one central object of the cluster
while the rest of 5 halos survive as 
separate entities inside the cluster
at the present epoch.

The thick solid curve in Figure  \ref{fig:mass_func_diff_z0}
presents the relation 
between the observed number density of $z=4$ LBGs and the predicted 
dark-halo mass of their descendants at $z=0$. 
This curve is derived by evolving the thick solid curve in 
Figure \ref{fig:mass_func_diff_z4}
to the present epoch using eq. (\ref{eq:bias0_sheth}). 
It is interesting to note that the data points for LBGs 
at $z=3$ and 5 are located close to this solid curve.
This may imply that descendants of LBGs 
at $z=3$, 4, and 5 are indistinguishable 
in terms of the number density and halo mass.

We find that 
the mass of descendants of dark halos hosting SCUBA sources
is about $4.5^{+2.6}_{-2.5}\times 10^{14} h_{70}^{-1} M_\odot$.
Thus the descendants of SCUBA sources will probably
be embedded in clusters of galaxies.
We then find that the data point of SCUBA sources 
is located near the high-mass end of the thick solid curve
in Figure \ref{fig:mass_func_diff_z0}. This would indicate
that SCUBA sources are very dusty and high
star-formation rate LBGs (a similar discussions
is found in \citealt{adelberger2000}).
The number density of SCUBA sources is about $1/10$ that 
of present-day normal galaxies in clusters.
This implies that the descendants of SCUBA sources are
a rare population in present-day cluster members.
Since the fraction of elliptical galaxies in present-day cluster
members is about 20\% \citep{dressler1980,whitmore1993}, the
descendants of SCUBA sources could be cluster ellipticals.
This inference supports the prediction by 
\cite{shu2001} based on the CDM model that descendants of
bright sub-mm sources should reside in present-day clusters
and that these objects are likely to be the progenitors of giant
ellipticals.

\section{Discussion}
\label{sec:discussions}

\subsection{Clustering Properties of LBGs at $z=4$ and $5$}
\label{sec:discussions_clustering}

\begin{figure}
   \plotone{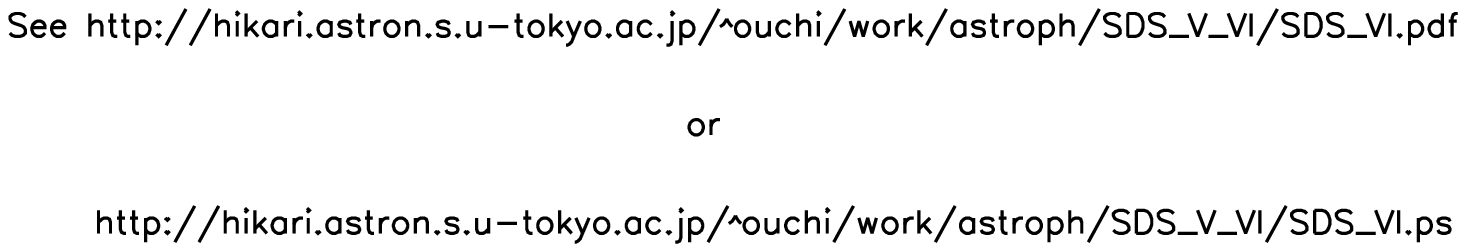}
   \caption{Correlation length $r_0$ (top panel)
   and galaxy-dark matter bias $b_g$ (bottom panel) as a function of redshift.
   The filled circles indicate our LBGs at $z=4-5$, and the 
   filled diamond is
   for LBGs at $z=3$ (\citealt{giavalisco2001}; 
   results based on the SPEC sample with $R<25.0$). These LBGs 
   at $z=3-5$ have a similar luminosity ($\gtrsim L^*$). The open circle 
   indicates LAEs at $z=4.9$ obtained by SDS II. 
   The filled pentagons, triangles, and squares
   are $r_0$ of galaxies at $z=0-1$, 
   derived by \citet{loveday1995}, \citet{carlberg2000}, and
   \citet{brunner2000}.
   The bias values of 
   these galaxies at $z=0-1$ are calculated in the same manner as 
   given in section \ref{sec:spatial_correlation}.
   The dotted line 
   shows the correlation length of
   the underlying dark matter 
   obtained with a non-linear theory \citep{peacock1996}.
   The dashed line indicates $r_0$ of dark matter predicted by
   linear theory 
   which we use in section \ref{sec:spatial_correlation}.
   The solid lines indicate $r_0$ and $b_g$ of galaxies with 
   $B<-19+5\log h$ predicted by the
   semi-analytic model of \citet{kauffmann1999}.
   Our results for LBGs at $z=3-5$ also show good agreement
   with the predictions of 
   the semi-analytic model for galaxies 
   with $m_R<25.5$ \citep{baugh1999}.   
   }
   \label{fig:z_r0_bias}
\end{figure}

\begin{figure}
   \plotone{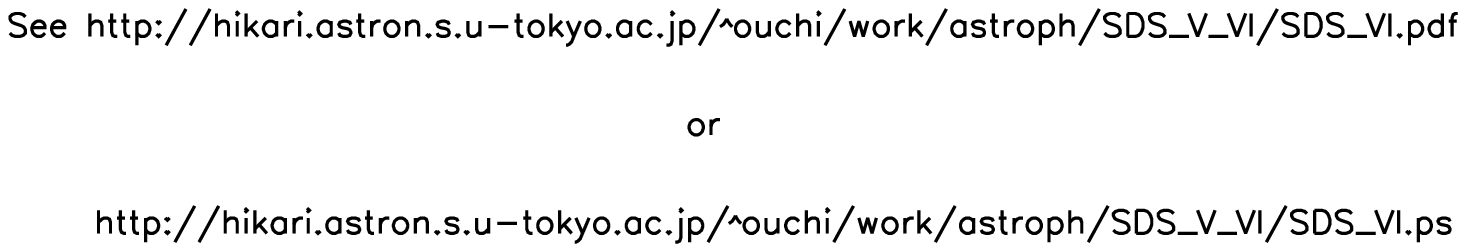}
   \caption{Magnitude dependence of the correlation length, $r_0$, 
   and the galaxy-dark matter bias, $b_g$, for LBGs at $z=4$ in the SDF.
   The four filled circles correspond to the four subsamples.
   The solid line shows the scaled Norberg's law. The ticks 
   in the top margin 
   indicate the number density of observed LBGs 
   calculated from the luminosity function
   given by SDS V. The ticks in the right margin
   indicate the mass of the hosting dark halos 
   calculated by the formula given by \citet{sheth1999} using
   $b_g$ we obtain (see text).}
   \label{fig:bias_norberg}
\end{figure}

Figure \ref{fig:z_r0_bias} shows
the observed correlation length, $r_0$ (in comoving units), and
galaxy-dark matter bias, $b_g$,
of LBGs at $z\simeq4$ and $5$ in our sample, 
together with those of galaxies 
at various redshifts between $z=0$ and $3$
taken from the literature.
The data points for LBGs (at $z=3-5$) are for galaxies with
$L\gtrsim L^*$ in UV continuum, while the other points 
at $0<z<1$ are for galaxies selected by optical continuum.
Figure \ref{fig:z_r0_bias} indicates that the clustering of LBGs
at $z=3-5$ is strongly biased against the underlying dark matter 
(dotted or dashed line).
The values for the bias parameter are
$b_g=3-5$ at $z=3-5$. The bias
increases monotonically with redshift. 
The increase in $b_g$ with redshift has been
predicted by semi-analytic models
\citep{baugh1999,kauffmann1999} and hydrodynamical simulations
\citep{blanton2000,yoshikawa2001}.
The observed $b_g$ values are consistent with
predictions by \citet{kauffmann1999} (solid line of Figure
\ref{fig:z_r0_bias}) for galaxies with $M<-19+5\log h$ in the $B$ band.
Although the selection criteria of galaxies are
different between ours (LBG selection)
and \citeauthor{kauffmann1999}'s ($B$-band selection),
their semi-analytic model
reproduces the observed trend of $b_g$ fairly well.

In Figure \ref{fig:z_r0_bias}, $r_0$ 
decreases with redshift over $z\sim 0-1$, 
has a minimum at $z\sim 1-3$, and then slightly increases at $z>3$.
This behavior can be explained by a combination of 
(i) the increase in bias ($b_g$) with redshift and 
(ii) the decrease in the correlation amplitude of dark matter 
($\xi_{\rm DM}$) with redshift, 
since $r_0$ is expressed as 
$r_0/(8 {\rm Mpc})=b_g^{(2/\gamma)} \xi_{\rm DM}({\rm r=8 Mpc})^{(1/\gamma)}$
(eq. \ref{eq:bias_detail}).
At low redshifts, the decrease in $\xi_{\rm DM}$ (with redshift)
dominates over 
the increase in $b_g$. This is the reason why $r_0$ decreases with redshift 
over $z\sim 0-1$.
On the other hand, the increase in $b_g$ is more effective 
at high redshifts, resulting in an increase in $r_0$ with redshift 
at $z>3$.
Note that these two factors roughly cancel with each other, 
and thus the change in $r_0$ over $z=0-5$ is modest 
compared with the change in $b_g$.
\citet{kauffmann1999} have found a qualitatively similar 
trend in the evolution of $r_0$ in their galaxy formation models. 
They have claimed that a dip in $r_0$ between $z=0$ and 1 occurs 
in $\Lambda$-dominated flat universes,
because structures form early and dark halos hosting 
$L^*$ galaxies are unbiased tracers of the dark matter 
over this redshift range.
To summarize, the observed behaviors of $b_g$ and $r_0$ 
found in this paper can be 
regarded as a piece of evidence supporting 
the biased galaxy formation scenario.

\subsection{Dark-Matter Properties of High Redshift Galaxies}
\label{sec:discussions_dark}

It is found from 
Figures \ref{fig:mass_func_diff_z4} and \ref{fig:mass_func_diff_z3} 
that the mass of dark halos 
hosting LBGs at $z=4-5$ ranges from $10^{11}$ to 
$10^{13} h_{70}^{-1}M_\odot$.
\citet{giavalisco2001},
\citet{bullock2002}, and \citet{moustakas2002} have 
estimated the mass of dark halos hosting LBGs at $z=3$ using 
analytic approaches based on the CDM model.
\citet{giavalisco2001} have found that the average mass of 
dark halos for LBGs with $R=23$, $25.5$, and $27$ is 
$\langle M\rangle =3.6$, $1.3$, and 
$0.6\times 10^{12} h_{70}^{-1}M_\odot$, respectively. 
\citet{bullock2002} have derived 
the minimum mass of dark halos, $M_{\rm min}$, and 
the mass of dark halos hosting one LBG, $M_1$, 
using the observed spatial correlation functions 
given by \citet{adelberger1998}, 
\citet{giavalisco2001}, and \citet{arnouts1999}. 
They have found 
$M_{\rm min} \simeq (0.6-1.1) \times 10^{10} h_{70}^{-1}M_\odot$
and $M_1=(0.9-1.4) \times 10^{13} h_{70}^{-1}M_\odot$.
\citet{moustakas2002} have obtained 
$M_{\rm min} \simeq 1.9 \times 10^{10} h_{70}^{-1}M_\odot$,
$M_1 \simeq 8.6 \times 10^{12} h_{70}^{-1}M_\odot$, and 
$\langle M\rangle \simeq 8 \times 10^{11} h_{70}^{-1}M_\odot$.
>From Figure \ref{fig:mass_func_diff_z3}, 
dark halos hosting LBGs at $z=3$
have $M_1 \simeq 10^{13}h_{70}^{-1}M_\odot$ , which is 
close to the values obtained by \citet{bullock2002}.
Similarly, the typical mass of dark halos hosting LBGs at $z=3$, 
$\langle M \rangle \sim 10^{12} h_{70}^{-1}M_\odot$, 
is close to those derived by \citet{giavalisco2001} and 
\citet{moustakas2002}. Thus, our results for LBGs at $z=3$
($M=1-10 \times 10^{12}$)
are consistent with those given by the previous works.
Furthermore, \citet{hamana2004} have derived recently 
the mass of dark halos of our LBGs and LAEs at $z=4$ and $5$
using the halo occupation function. They directly 
fit predicted angular correlation functions and 
number density to data for LBGs and LAEs 
(obtained in this paper and SDS II), 
and obtain $\langle M\rangle$,
$M_1$, and $M_{\rm min}$. They assume the power-law
occupation number function
and the same analytic halo models 
(\citealt{sheth1999}) as used in this paper.
They obtain $\langle M\rangle = 9\times10^{11}$, $6\times10^{11}$,
and $1\times 10^{12} h_{70}^{-1}M_\odot$ for LBGs at $z=4$ and $5$
and LAEs at $z=5$, respectively, 
which are comparable to our results
($M=4\times10^{11}$, $9\times 10^{11}$, and $1\times 10^{12}$).
\citet{hamana2004} have found an indication that
the simple halo model may not work for small-scale clustering
($\theta < 120''$ corresponding to $r<3.1 h_{100}^{-1}$ Mpc) of LAEs.
In our analysis, we obtain the bias at a large scale 
(at $8 h_{100}^{-1}$ Mpc), and do not examine whether
$\xi_{\rm DH}$ is fit to $\xi{\rm (LAE)}$ on small scales.
In this sense, our analysis is simpler than theirs.

In Table \ref{tab:halo_mass_lstar},
we summarize the properties of dark halo
hosting $L\gtrsim L^*$ LBGs (and LAEs), 
which are extracted from Table \ref{tab:halo_mass}.
We find the typical mass of dark halos 
hosting $L\gtrsim L^*$ LBGs is 
$1.3^{+1.0}_{-0.6}\times 10^{12} h_{70}^{-1}M_\odot$,
$1.0^{+0.9}_{-0.7}\times 10^{12} h_{70}^{-1}M_\odot$, and
$8.9^{+8.9}_{-6.4}\times 10^{11} h_{70}^{-1}M_\odot$
for $z=3$, $4$, and $5$, respectively.
Thus the dark halo masses hosting $L\gtrsim L^*$ LBGs
are almost constant around $1\times 10^{12} h_{70}^{-1}M_\odot$
which is comparable to 
the total mass of the Milky Way Galaxy at the
present epoch. Although there might 
be a trend that the typical mass decreases slightly with 
redshift, we find no clear difference in the mass range 
of dark halos hosting LBGs with $L>L^*$ 
($M\lesssim-20.5$ or, equivalently, 
SFR$\gtrsim 10$ [$40$]$ h_{70}^{-2} M_\odot$yr$^{-1}$
before [after] extinction correction) over $z=3-5$.
This finding suggests that the mass-to-luminosity ratio of 
LBGs (the mass of a hosting halo divided by a UV-luminosity 
of the LBG in the halo) does not change largely from $z=3$ to 5. 
In other words, the star formation efficiency is almost constant 
over $z=3-5$.
This suggestion may not be consistent with the theoretical prediction 
that the star-formation efficiency increases with redshift 
(e.g., \citealt{hernquist2003}).

Figure \ref{fig:mass_func_diff_z4} 
and Table \ref{tab:halo_mass} show that the occupation number, 
$N_{\rm occup}$, of LBGs for 
massive dark halos ($\sim 10^{13} h_{70}^{-1}M_\odot$) is 
close to unity while $N_{\rm occup}$ for less massive halos 
($\sim 10^{12} h_{70}^{-1}M_\odot$) is about 0.1.
Thus, it is concluded that more massive halos possess 
a larger number of LBGs.
Note that a similar feature is found for the 2dFGRS galaxies at $z=0$ 
in the range of dark halo mass of of $10^{12} - 10^{13}h_{70}^{-1}M_\odot$ 
as shown in Figure \ref{fig:mass_func_diff_z0}.

\subsection{Unified View for Various Kinds of High-$z$ Galaxies}
\label{sec:discussions_unified}

In this section, we discuss relations between LBGs, LAEs, 
$K$-band selected galaxies (FIRES), and SCUBA sources.
These galaxies are named after the search techniques 
(section \ref{sec:introduction}), 
and are all located at high redshifts.
Little has been known to date about relations between these galaxy 
populations.

Figure \ref{fig:mass_func_diff_z0} plots 
the observed number densities of these high-$z$ galaxies 
against the predicted masses of dark halos of
their descendants at $z=0$.
In this figure, the thick solid curve 
corresponds to the best-fit function 
to the observed relation between the bias parameter and 
the number density for our $z=4$ LBGs (Figure \ref{fig:bias_norberg} 
of section \ref{sec:spatial_segregation_lbg_uv}).
Interestingly, the data points of LBGs at both $z=3$ and $z=5$ 
(blue and red circles) are located very close to this curve.
This suggests that LBGs at $z=3-5$ are a single population 
in terms of the mass of hosting halos and the halo occupation 
number.

We propose here that LBGs at $z=3-5$ come from a single parent 
population of galaxies, 
assuming that any dark halo more massive than a critical value 
has such parent galaxies and that the stellar mass of these galaxies
is proportional to the mass of the dark halo.
These parent galaxies can be detected by NIR surveys such as 
FIRES, because NIR surveys sample galaxies according to their 
rest-frame optical luminosity, i.e., their stellar mass.
Figure \ref{fig:mass_func_diff_z3} shows 
that the number density of FIRES galaxies is 
10 times higher than that of LBGs at the same mass range 
of dark halos.
If LBGs are star-forming galaxies with intermittent 
star formation activities as suggested 
in section \ref{sec:galaxy_hosting_z4}, 
this difference in the number density reflects probably 
the total duration time of active star formation in galaxies.
Both LBGs and FIRES galaxies are seen at $z=2.5$ and $3.5$, 
and the cosmic time between these two redshifts is 
$8.0 \times 10^8$ yrs.
If we assume that LBGs are visible in optical observations 
(i.e., bright in the rest-frame UV wavelength) only when 
they are in a phase of active star formation, 
then the total duration time of star formation between $z=2.5$ and $3.5$ 
is estimated to be $8.0 \times 10^8 / 10 = 8 \times 10^7$ yrs.
This value is comparable to the typical age of stellar populations 
of LBGs, $7 \times 10^7$ yrs, given by \citet{papovich2001}.
They have derived the age from a fitting of 
model spectra to the observed optical-to-NIR colors of LBGs 
at $z\sim 3$, assuming that the star formation history of LBGs 
is approximated by an exponentially decaying star-formation rate.
Furthermore, if we assume that the typical star-formation rate
of LBGs is $20-200M_\odot/$yr 
(whose values are extinction-corrected by a factor of 4: 
see Figure 16 and section 5 of SDS V),
then the stellar mass accumulating 
from $z=5$ to 3 ($\Delta t = 9.6\times 10^{8}$yrs) is estimated 
to be roughly $9.6 \times 10^8 [{\rm yr}]/10 \times (20-200) 
[M_\odot {\rm yr}^{-1}] \sim (2-20) \times 10^9 M_\odot$.
This value is comparable to the typical stellar masses of LBGs at $z\sim 3$ 
derived by \citet{shapley2001} and \cite{papovich2001}.
%
%
%
On the other hand, 
if LBGs are FIRES galaxies in a phase of star formation, 
LBGs should have similar stellar masses to FIRES galaxies.
This prediction is supported by 
\citet{franx2003}'s finding that while FIRES galaxies
are slightly less luminous in the observed-frame $K$ band than
the brightest LBGs, their stellar masses 
are comparable to, or higher than, the brightest LBGs'. 
%
%

%
%
The data point of SCUBA sources (star mark) is located at the 
high-mass end of the solid curve in Figure \ref{fig:mass_func_diff_z0}. 
This coincidence would indicate that SCUBA sources and LBGs 
belong to the same population.
As discussed in section \ref{sec:spatial_segregation_lbg_uv},
brighter LBGs have larger dark-halo masses, 
indicating that LBGs with higher star-formation rates 
reside in more massive dark halos.
The star formation rates of SCUBA sources, $\sim 10^3 M_\odot /$yr, 
are (much) higher than those of LBGs, $20-200 M_\odot /$yr
after extinction correction.
Thus, the correlation found for our LBGs at $z=4$ 
holds for SCUBA sources as well.
We have also found a possible connection of dusty LBGs to SCUBA sources
in the analysis of section \ref{sec:spatial_segregation_lbg_dust}.

We have found a probable connection between LBGs, FIRES galaxies, 
and SCUBA sources as discussed above. 
However, LAEs at $z=4.9$ cannot be understood in a similar way.
Since LAEs are also star-forming galaxies, they should be located 
near the solid curve in Figure \ref{fig:mass_func_diff_z0}, 
if they belong to the same population as LBGs.
%
%
%
In this case, LAEs are expected to be located 
near the low-mass end of the solid curve, 
because the typical star-formation rate of LAEs 
is only about $10$ ($3$) $M_\odot /$yr with (without) 
dust extinction correction. 
\footnote{
We estimate these SFR values 
from their median UV continuum flux. 
If we assume LAEs to be dust free, we obtain the typical 
SFR of about 3$M_\odot /$yr.
On the other hand, if we assume that the dust extinction of LAEs 
is comparable to that of LBGs at z=4, i.e., E(B-V)=0.15,
we obtain 10$M_\odot /$yr for the extinction corrected SFR.
These values are comparable to, or lower than, 
the average extinction-corrected SFR for 
the faintest subsample of $BRi$-LBGs, 
although the dust extinction amount of LAEs is not yet 
well constrained.
However, as found from Figure \ref{fig:mass_func_diff_z0}, 
LAEs deviate largely from the solid curve toward higher halo masses;
dark halos hosting LAEs are as massive as those hosting bright LBGs.
}
%
%
%
We think that this apparent discrepancy is related to
(i) a large cosmic variance in the LAE sample of SDS II and/or 
(ii) a selection bias of LAEs.
First, as we note in section \ref{sec:spatial_angular},
the surveyed volume for their LAEs is
just $9.0\times 10^4 h_{70}^{-3}$Mpc$^{3}$, and thus
the results could be strongly affected by the field variance.
In reality, \citet{shimasaku2003} have found that their LAEs
make a large-scale structure, and it is possible that
the clustering derived in SDS II is biased high 
due to the large-scale structure.
Second, star-forming galaxies with relatively bright Lyman $\alpha$ emissions 
can be selected as LAEs even when their star formation rates 
(i.e., UV luminosities) are considerably low.
We have found in SDS II
that LAEs brighter in Lyman $\alpha$ 
emission are clustered more strongly while clustering strength 
does not clearly correlate with UV luminosity.
If this correlation holds for all star-forming galaxies, 
the clustering strength of LAEs should be higher than 
that of general star-forming galaxies with similar star-formation 
rates, resulting in very massive dark halos for LAEs.
The cause of this 'Lyman $\alpha$ luminosity bias' is 
not clear to us, but a possible explanation is that 
the Lyman $\alpha$ luminosities of galaxies increase 
(only) when they infall toward massive dark halos.
When galaxies infall toward massive halos, star formation may be 
triggered in them due to, for instance, interactions with 
the halos. 
Since the duration of strong Lyman $\alpha$ emission 
is very short, galaxies with strong Lyman $\alpha$ emission 
will be found only around massive halos, 
and thus LAEs have strong clustering.
\citet{scannapieco2003} have discussed a similar scenario 
for high-$z$ starburst galaxies.
%
%
%
In this explanation, we assume that Lyman $\alpha$ luminosity 
correlates with star formation rate fairly well.
However, this assumption may not necessarily be true.
For instance, \citet{shapley2001} find that LBGs with
stronger Lyman $\alpha$ emissions have older stellar populations, 
contrary to the predictions from simple H{\sc ii} region models. 
In any case, our knowledge about the nature of LAEs is still poor.
It is of great importance to accumulate the data of LAEs 
by new observations 
to infer more accurately their fundamental quantities 
such as halo mass, stellar mass, and star formation rate.
%
%
%

%
%
%
Finally, we summarize our unified view for high-$z$ 
galaxies (i.e., LBGs, FIRES galaxies, SCUBA sources, and LAEs), 
which we have proposed in this subsection. 
At $z \sim 3-5$, dark halos with $\sim 10^{12} h_{70}^{-1}M_\odot$ 
host a few galaxies (section \ref{sec:galaxy_hosting_z3}) 
whose stellar masses are large enough 
to be detected as FIRES galaxies.
Most of the galaxies hosted by these dark halos 
are quiescent in star-formation, 
but about one tenth have a moderately strong star-formation 
activity (SFR$\sim 20-200M_\odot/$yr),
producing a bright UV continuum and thus being identified as LBGs.
Hence, FIRES galaxies are the parent population for LBGs 
according to our view.
LBGs found at $z\sim 3,4,$ and 5 are not necessarily 
on the same evolutionary track, 
since the duration of a single star formation for LBGs 
is expected to be much shorter than the cosmic times 
between $z=3$, 4, and 5.
More massive dark halos tend to host LBGs with higher SFRs,
and extremely massive dark halos 
with $\sim 10^{13} h_{70}^{-1}M_\odot$ have
a starburst galaxy with SFR$\sim 10^3M_\odot/$yr, 
which is identified as a SCUBA source. 
The relationship between LAEs and the other high-$z$ galaxies 
is not clear. 
%
%
%

\subsection{Formation History of Galaxies and Large-Scale Structures
Based on the Observational Results}
\label{sec:discussions_formation}

In sections \ref{sec:discussions_unified}, 
we have proposed that
LBGs and SCUBA sources are high-$z$ star-forming galaxies
emerged from the same parent population. 
In our proposal, high-$z$ galaxies experience 
many episodic star-forming phases with star-formation rate of
$20-200 h_{70}^{-2} M_\odot$ yr$^{-1}$,
whose characteristic total duration time 
is $\sim 80$Myr (per 1 Gyr). The diversity
of the properties of star-formation galaxies may be due 
to the variety of star-formation rate.
These galaxies accumulate stars through episodic star-formation
activities of $20-200 h_{70}^{-2} M_\odot$ yr$^{-1}$,
and have stellar masses of $\sim 1\times 10^{10}M_\odot$
at $z=3$. 

In SDS V, we have found 
that the luminosity function of LBGs
does not change from $z=4$ to $z=3$, while the number of bright LBGs
($M_{\rm 1700}\lesssim -22$; extinction-corrected 
SFR is $\gtrsim 100 h_{70}^{-2} M_\odot$ yr$^{-1}$) 
drops toward $z=5$. If we assume that the star-formation efficiency
is constant from $z=5$ to $z=4$, this implies that
massive galaxies (with large cold gas reservoirs) are rare at $z\gtrsim 5$
since matter (baryon + dark matter) has not been assembled by the
gravitational instability at $z=5$. However, the cosmic star-formation
rate density is almost constant 
from $z\simeq 5$ to $z\simeq 1$ (SDS V). 
This is because
the major contributors to the cosmic star formation are galaxies
with smaller star-formation rates $\sim 1 h_{70}^{-2} M_\odot$yr$^{-1}$
(SDS V).
Then, most of the galaxies with 
active star formation of $\sim 100 h_{70}^{-2} M_\odot$ yr$^{-1}$
disappear in the present-day
Universe (Figure 16 of SDS V).
This reduction of galaxies with 
intensive star-forming activities leads to the
decrease in the star-formation rate density from $z\simeq 1$
to $z=0$.

On the other hand, dark matter is continuously assembled 
through the cosmic time (Figure \ref{fig:z_r0_bias}).
Thus the clustering amplitude, or correlation length,
of the dark matter monotonically increases with time.
However, the correlation length of galaxies
with $L\gtrsim L^*$
does not significantly change from $z=5$ to $z=0$
because the galaxy-dark matter bias ($b_g$) decreases
with time.
Thus, the galaxy clustering at $z=5$ is similar
to that at the present epoch.
This is the reason why 
SDS IV discovered the large-scale structure of galaxies
at $z=5$.
The size of the large-scale structure at $z=5$
is comparable to those in the present-day Universe
found by, e.g., \citet{geller1989}.

We have found that dark halos hosting high-$z$ galaxies (LBGs, LAEs etc.)
are considerably massive, $\sim 10^{12}h_{70}^{-1}M_\odot$,
which is comparable to the total mass of the Milky Way Galaxy.
The values of the bias parameter for such massive halos are 
very high at the observed redshifts. 
The large masses of dark halos hosting high-$z$ galaxies
imply that they will have evolved into groups and clusters
of galaxies at the present epoch
(section \ref{sec:galaxy_descendants}
and Figure \ref{fig:mass_func_diff_z0})
in a statistical sense.
This would indicate that the majority of present-day field galaxies
were formed at $z\lesssim 3$, or that high-$z$ progenitors of
field galaxies are much fainter than the limiting magnitudes
of today's observations ($i'\sim 27$ and $K\sim 24$).

\section{Conclusions}
\label{sec:conclusions}

We detect 2,600 Lyman break galaxies (LBGs) at $z=3.5-5.2$ 
in the deep ($i'\simeq 27$) and wide-field (1,200 arcmin$^2$)
data in the
Subaru Deep Field (SDF) and the Subaru/XMM Deep Field (SXDF).
First, we derive clustering properties 
(angular correlation function and correlation length etc.)
of LBGs at $z=4$ and $5$ and present the observational results.
We then combine these observational results with
the Cold Dark Matter model to understand
the formation and evolution of galaxies (luminous matter) and
dark halos (dark matter)
simultaneously and obtain theoretical implications.
The major findings of our study are categorized in
(a) observational results and (b) theoretical implications
as follows.\\

(a) Observational results:

1. We calculate angular two-point correlation functions of
LBGs, and find clear signals
for LBGs at $z =4$ and $5$. 
We estimate the correlation
length, $r_0$, of spatial two-point correlation function ($\xi_g$)
by Limber deprojection using
the redshift distribution obtained by 
simulations (section \ref{sec:data}).
We obtain $r_0=4.1\pm0.2 h_{100}^{-1}$ Mpc and
$5.9^{+1.3}_{-1.7}h_{100}^{-1}$ Mpc
for LBGs at $z=4$ 
and $z=5$, respectively.
We also calculate $r_0$ for bright galaxies
($L\gtrsim L^*$, i.e., 
$i'<25.3$ for LBGs at $z=4$ and $z'<25.8$ for LBGs at $z=5$)
to obtain $r_0= 5.1^{+1.0}_{-1.1} h_{100}^{-1}$ Mpc 
and $5.9^{+1.3}_{-1.7} h_{100}^{-1}$ Mpc
for LBGs at $z=4$ and $5$,
and compare them with that of $z=3$ LBGs with similar luminosities,
to find that the correlation length is
almost constant at $\sim 5h_{100}^{-1}$Mpc
from $z=3$ to $z=5$.

2. We find that the correlation amplitude of galaxies is larger than
that of underlying dark-matter ($\xi_{\rm DM}$) at high redshifts 
(at least at $z\gtrsim3$).
We calculate the galaxy-dark matter bias, 
$b_g\equiv \sqrt{\xi_g/\xi_{\rm DM}}$ at $8h_{100}^{-1}$Mpc,
for LBGs.
We find that $b_g$ of $L\gtrsim L^*$ LBGs at $z=4$ and $z=5$
is $3.5^{+0.6}_{-0.7}$ and $4.6^{+0.9}_{-1.2}$, implying that
the distribution of LBGs at $z\gtrsim 3$ is highly biased against the 
underlying dark matter. 
The bias of galaxies brighter than $\simeq L^*$
monotonically increases with redshift, supporting
the biased-galaxy formation scenario.

3. We make sub-samples of LBGs at $z=4$ binned according to magnitude,
and calculate the correlation length for each sub-sample.
We find that a brighter sub-sample has a larger correlation length
(or larger galaxy-dark matter bias): 
the brightest sub-sample ($M\simeq M^*-1$) has 
$r_0=7.9^{+2.1}_{-2.7} h_{100}^{-1}$ Mpc 
while the faintest sub-sample ($M\simeq M^*+0.5$) 
has $r_0=2.4^{+0.2}_{-0.5} h_{100}^{-1}$ Mpc.
Similarly, we make subsamples of LBGs at $z=4$
according to $E(B-V)$ estimated from the UV-continuum
color, $i'-z'$. 
We find that the dustiest LBGs in our sample ($E(B-V)\sim 0.4$) have 
$r_0=8.2^{+2.9}_{-4.2} h_{100}^{-1}$Mpc, which is marginally 
larger than the average. These dependences of $r_0$ on
UV luminosity and dust extinction
seen in LBGs are in accord with the fact that SCUBA sources,
which are thought to be very dusty starburst galaxies, have a
large correlation length, $r_0\sim 10h_{100}^{-1}$ Mpc.

\begin{figure}
   \plotone{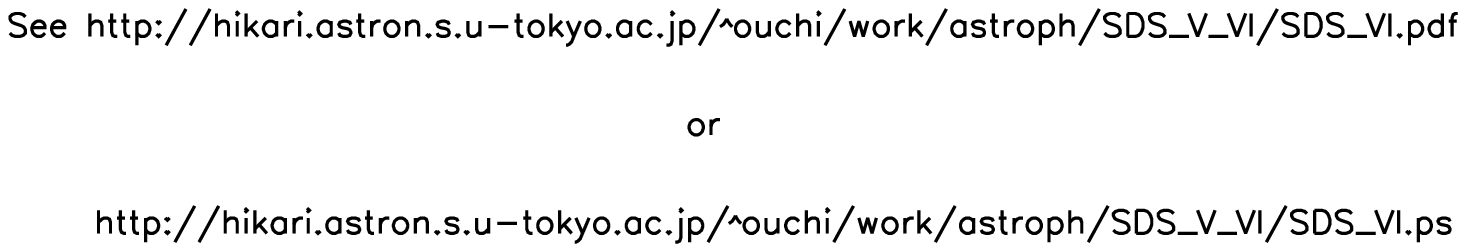}
   \caption{Mass function of dark halos hosting LBGs
   at $z=4$.
   The ordinate shows either the number density of LBGs 
   in a given dark-halo mass range (filled circles), or
   the number density of dark halos (thin solid line).
   $\Delta M$ shown in the ordinate axis means the number density in the
   mass range of 1 magnitude, i.e., $\Delta M= M/(2.5\log(e)) = 0.92M$.
   The thin solid line
   indicates the mass function of all dark halos at $z=4$
   calculated using the formula given by \citet{sheth1999}.
   The four filled circles correspond
   to the four data points shown in Figure \ref{fig:bias_norberg}. 
   The thick solid 
   line corresponds to a fit of the scaled Norberg's law to the four
   data points of Figure \ref{fig:bias_norberg}.
   }
   \label{fig:mass_func_diff_z4}
\end{figure}

\begin{figure}
   \plotone{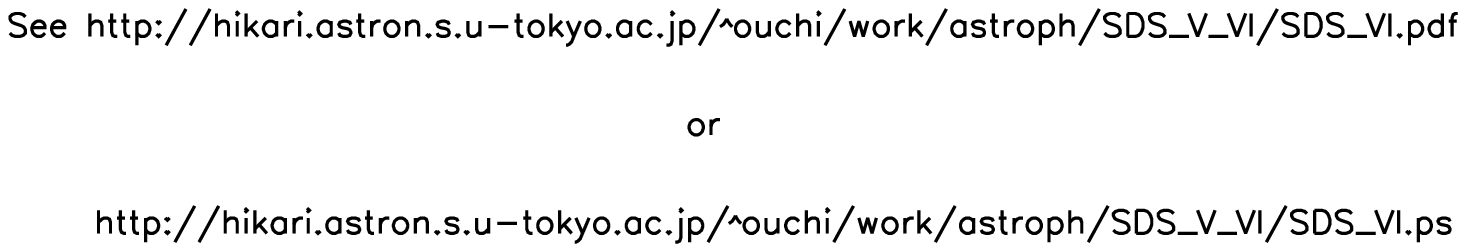}
   \caption{Same as Figure \ref{fig:mass_func_diff_z4}, but for
   galaxies at $z=3$.
   The two filled circles show LBGs at $z=3$ in the SPEC sample (right)
   and the PHOT sample (left) of \citet{giavalisco2001}.
   The open circle with an arrow indicates the measurement for 
   the HDF-N sample of \citet{giavalisco2001}. 
   The filled triangle corresponds to all galaxies found
   in FIRES \citep{daddi2003} which is a $K$-band limited 
   high-$z$ galaxy survey with $K<24$ \citep{labbe2003}, 
   while the two open triangles are for FIRES galaxies with
   blue color (left; $J-K<1.7$) and red color (right: $J-K<1.7$),
   respectively. 
   The star corresponds to SCUBA sources whose correlation length
   is reported to be 
   $r_0=12.8\pm4.5 h_{100}^{-1}$ Mpc by \citet{webb2003b}
   (see \citealt{scott2002} for their number density).
   }
   \label{fig:mass_func_diff_z3}
\end{figure}

\begin{figure}
   \plotone{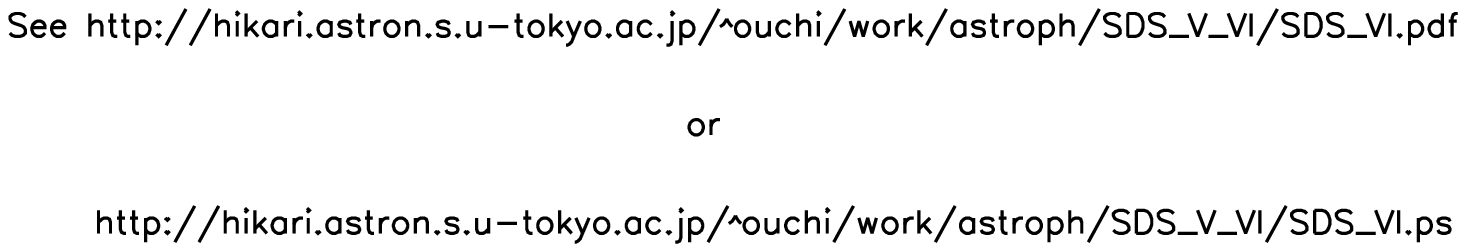}
   \caption{Same as Figure \ref{fig:mass_func_diff_z4}, but for
   LBGs and LAEs at $z=5$.
   The filled circle and filled square indicate LBGs and LAEs,
   respectively. The open square is also for LAEs, 
   but when a higher dark-halo
   mass, $3.5\times 10^{12}h_{70}^{-1} M_\odot$, is adopted.
   See footnote of section \ref{sec:galaxy_hosting_z5} for details.
   }
   \label{fig:mass_func_diff_z5}
\end{figure}

\begin{figure}
   \plotone{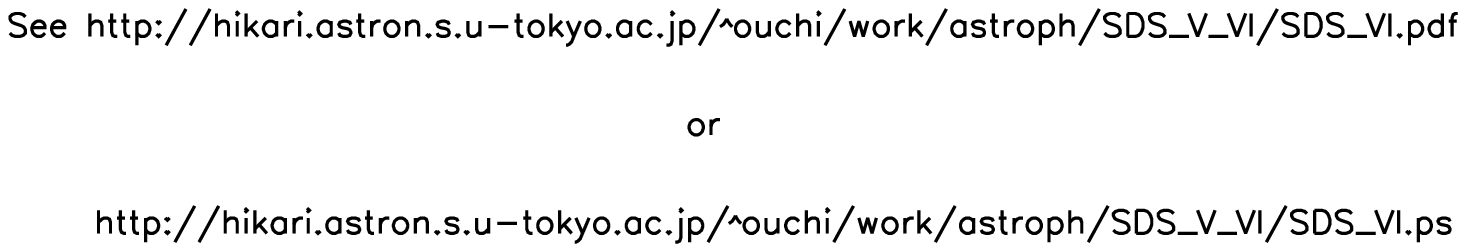}
   \caption{Same as Figure \ref{fig:mass_func_diff_z4}, but for
   present-day descendants of high-$z$ galaxies at $z=3$, $4$, and $5$.
   We calculate the dark halo mass of descendants using \citet{sheth2001},
   but assign to the descendants the same number densities
   as measured for galaxies at high redshifts.
   The blue, black, and pink filled circles show descendants of LBGs 
   at $z=3$, $4$, $5$, respectively. The red square is for descendants
   of LAEs at $z=5$. The blue triangle and star mark denote all
   $K$-selected galaxies with $K<24$ (FIRES) 
   and submillimeter-selected galaxies
   with $>3$mJy (SCUBA). The thick solid line corresponds to the scaled
   Norberg's law at $z=4$ (Figure \ref{fig:mass_func_diff_z4})
   evolved to $z=0$ assuming the model of \citet{sheth2001}. 
   The dotted line indicates the relation for 
   2dFGRS galaxies with $M_{b_g}\lesssim -19$,
   which is calculated from the mass function of dark halos (solid line)
   and the occupation number derived by \citet{vandenbosch2003}
   }
   \label{fig:mass_func_diff_z0}
\end{figure}

(b) Theoretical implications: 

4. We investigate properties of dark halos hosting
LBGs based on 
the luminosity function (SDS V)
and correlation function (section \ref{sec:spatial}) of our LBGs,
using the Cold Dark Matter model \citep{sheth1999,sheth2001}.
We estimate the mass of dark halos hosting LBGs at $z=4$
from the the galaxy-dark matter bias at a large separation,
$b_g$, and the number density of galaxies from the luminosity function.
We find that 
LBGs at $z=4$ reside in dark halos whose mass ranges over
$10^{11}-10^{13} h_{70}^{-1}M_\odot$,
and that bright LBGs ($M\simeq M^*-1$) 
have more massive dark halos 
($4.6^{+4.0}_{-3.4}\times 10^{12} h_{70}^{-1}M_\odot$)
than faint LBGs 
($M\simeq M^*+0.5$; $1.7^{+1.1}_{-0.7}\times 10^{11} h_{70}^{-1}M_\odot$).
We compare the number density of LBGs with that of
all dark halos existing at $z=4$ (Figure \ref{fig:mass_func_diff_z4})
to find that the number density of massive dark halos
($\simeq 10^{13} h_{70}^{-1}M_\odot$) is
comparable to that of LBGs while the number density 
of less massive dark halos 
($\lesssim 10^{12} h_{70}^{-1}M_\odot$) is 3 to 10 times
higher than that of LBGs. This indicates that 
the majority of the low-mass 
($\lesssim 10^{12} h_{70}^{-1} M_\odot$) dark halos have no LBGs,
and that most of the galaxies at $z=4$ do not have
UV-continuum emission originated from star-formation
activity which is strong enough to exceed our detection limit
($M_{1700}\sim -20$ or 
$\sim 20 h_{70}^{-2} M_\odot$yr$^{-1}$ for extinction-corrected 
star-formation rate).
Thus, we propose that LBGs are high-$z$ galaxies which
have episodic star formation and are detected only when star-formation
activity is high.

5. We perform a similar analysis to LBGs at $z=5$ in our data
and LAEs at $z=5$ obtained by SDS II.
We find that both LBGs and LAEs at $z=5$ have dark halos with
$\sim 10^{12}h_{70}^{-1}M_\odot$. The number density of
LBGs is about 2 times smaller than that of dark halos
with $\sim 10^{12}h_{70}^{-1}M_\odot$. 
On the other hand, the number density
of LAEs is about 3 times higher than that of dark halos.
Thus, dark halos of $\sim 10^{12}h_{70}^{-1}M_\odot$
preferably host multiple LAEs.

6. We apply a similar analysis to LBGs, $K$-band selected
galaxies (FIRES galaxies), and SCUBA sources at $z=3$ 
whose correlation lengths and number densities are
given by \citet{giavalisco2001}, \citet{daddi2003},
and \citet{webb2003b}, respectively.
We find that LBGs with $L>L^*$ and FIRES galaxies have
dark halos whose mass is $\sim 10^{12}h_{70}^{-1}$M$_\odot$,
while dark halos for SCUBA sources are 
more massive ($\sim 3\times 10^{13}h_{70}^{-1}$M$_\odot$),
suggesting that SCUBA sources could reside in proto-cluster regions.
The number density of LBGs
is three times lower than 
that of dark halos existing at $z=3$,
while the number density of FIRES galaxies is about 4 times higher
than that of dark halos. This implies that
only about $1/10$ ($\simeq 0.3/4$) 
of galaxies residing in 
dark halos with $\sim 10^{12} h_{70}^{-1} M_\odot$
have an active-star formation whose star-formation rate (SFR) exceeds
$\sim 5 h_{70}^{-2} M_\odot$yr$^{-1}$ 
($\sim 20 h_{70}^{-2} M_\odot$yr$^{-1}$ if 
dust extinction is corrected),
so that they are identified as LBGs by
their bright UV continuum. 
This difference between LBGs and FIRES galaxies
may support the idea that LBGs are galaxies which happen 
to be bright at UV wavelength due to episodic star formation.

7. We find that the typical mass of dark halos 
hosting $L\gtrsim L^*$ LBGs is about
$1\times 10^{12} h_{70}^{-1}M_\odot$, 
which is comparable to the total mass of the Milky Way Galaxy at the
present epoch, and that the typical mass is
almost constant over $z=3$ to $5$.
It implies that the mass-to-luminosity ratio of 
LBGs (the mass of a hosting halo divided by the UV-luminosity 
of a LBG in the halo) does not change largely from $z=3$ to 5  
or, equivalently, the star formation efficiency is almost constant 
over $z=3-5$.

8. Using the CDM model, we estimate the mass of present-day
dark halos hosting descendants of high-$z$ ($z=3-5$) galaxies.
We find that the mass of the descendants ranges
from 
$10^{13}h_{70}^{-1}M_\odot$ to 
$10^{15}h_{70}^{-1}M_\odot$, 
which are comparable to the mass of present-day clusters and groups.
Thus, dark halos hosting the observed high-$z$ galaxies 
will evolve into clusters and groups. In other words, 
most of the descendants of high-$z$ galaxies studied here are
probably member galaxies in clusters and groups today.
The number densities of FIRES galaxies, LAEs, and faint LBGs
exceed that of present-day galaxies with $M_{b_g}\lesssim -19$.
If most of these high-$z$ galaxies
are progenitors of present-day galaxies 
with $M_{b_g}\lesssim -19$, they should experience
mergers at least a few times from $z=3-5$ to $z=0$. \\


\acknowledgments
We would like to thank the Subaru Telescope staff
for their invaluable help in commissioning the Suprime-Cam
that made these difficult observations possible.
%
%
%
We thank the anonymous referee for 
suggestions and comments
that improved this article.
%
%
%
M. Ouchi acknowledges
support from the Japan Society for the
Promotion of Science (JSPS) through JSPS Research Fellowships
for Young Scientists.



%
%



\clearpage 

\clearpage 

\clearpage 

\clearpage 

\clearpage 

\clearpage 

\clearpage 

\clearpage

\clearpage

\clearpage

\clearpage

\clearpage

\clearpage


\clearpage

\begin{deluxetable}{cccrl}
\tabletypesize{\scriptsize}
\tablecaption{Photometric Samples of Galaxies\label{tab:sample}}
\tablewidth{0pt}
\tablehead{
\colhead{Field Name} & 
\colhead{Sample Name} &
\colhead{Detection Band} & 
\colhead{Number} &
\colhead{Magnitude Limit\tablenotemark{\ddag}} 
}
\startdata
SDF  & $BRi$-LBG & $i'$ & 1,438 & $i'<26.3$\\
SDF  & $Viz$-LBG & $z'$ &   246 & $z'<25.8$\\
SDF  & $Riz$-LBG & $z'$ &    68 & $z'<25.8$\\
SXDF & $BRi$-LBG & $i'$ &   732 & $i'<25.8$\\
SXDF & $Viz$-LBG & $z'$ &   34\tablenotemark{\dagger} & $i'<25.3$\\
SXDF & $Riz$-LBG & $z'$ &    38 & $i'<25.3$
\enddata

\tablenotetext{\ddag}{Total magnitudes}
\tablenotetext{\dagger}{$Viz$-LBGs of SXDF are selected by 
eq. (\ref{eq:lbgselection_SXDFVizLBG}).
}

\end{deluxetable}

\clearpage

\begin{deluxetable}{ccrr}
\tabletypesize{\scriptsize}
\tablecaption{Objects selected by two criteria\label{tab:detection_coincide}}
\tablewidth{0pt}
\tablehead{
\colhead{Field Name} & 
\colhead{Criteria\tablenotemark{\dagger}} &
\colhead{$N_{\rm OBS}$\tablenotemark{\dagger\dagger}} & 
\colhead{$N_{\rm EST}$\tablenotemark{\P}}
}
\startdata
SDF  & $Viz$+$Riz$ & 37 & $41\pm8.2$ \\
SDF  & $Viz$+LAE\tablenotemark{\ddag}   &  6 & $4\pm0.4$ \\
SDF  & $Riz$+LAE\tablenotemark{\ddag}   &  2 & $2\pm0.3$ \\
SXDF & $Viz$+$Riz$ &  3 & $5\pm3.1$ 
\enddata
\tablenotetext{\dagger}{Two selection criteria.}
\tablenotetext{\dagger\dagger}{Number of objects identified 
by both selection criteria.}
\tablenotetext{\P}{Number of objects estimated from
the contamination, completeness, and redshift distribution 
of the samples
as described in the text.}
\tablenotetext{\ddag}{LAE means the criteria for selection of
Lyman $\alpha$ emitters (LAEs)
at $z=4.86\pm0.03$ in the SDF. Details of the LAE sample
are described in SDS II.}

\end{deluxetable}

\clearpage

\begin{deluxetable}{cccrrrrr}
\tabletypesize{\scriptsize}
\tablecaption{Summary of the clustering measurements.\label{tab:acorr_comb}}
\tablewidth{0pt}
\setlength{\tabcolsep}{0.02in}
\tablehead{
\colhead{Sample Name} &
\colhead{$\langle z \rangle$} & 
\colhead{Criteria\tablenotemark{\dagger}} & 
\colhead{$N$\tablenotemark{\dagger\dagger}} &
\colhead{$f_c$(\%)\tablenotemark{\ddag}} &
\colhead{$\beta$} &
\colhead{$A_\omega^{\rm raw}$} &
\colhead{$A_\omega$\tablenotemark{\ddag\ddag}}
}
\startdata
$BRi$-LBG & 4.0 & all ($i'\lesssim26$) & 1438+732 &  2 & $0.90^{+0.11}_{-0.07}$ & $2.3^{+0.9}_{-0.6}$ & $2.4^{+0.9}_{-0.6}$\\
\tableline
$BRi$-LBG & 4.0 & all ($i'\lesssim26$) & 1438+732 &  2 & $0.8$(fix) & $1.7^{+0.1}_{-0.2}$ & $1.7^{+0.2}_{-0.1}$ \\
$Viz$-LBG & 4.7 & all ($z'<25.8$)\tablenotemark{\P} &  246 & 26 & $0.8$(fix) & $2.1^{+0.9}_{-0.9}$ & $3.8^{+1.7}_{-1.7}$\\
$Riz$-LBG & 4.8 & all ($z'<25.8$) &   68 & 40 & $0.8$(fix) & $0.4^{+3.2}_{-3.0}$ & $1.2^{+8.9}_{-8.4}$\\
\tableline
$BRi$-LBG & 4.0 & $i'<25.3$\tablenotemark{\P} &  357 & 1 & $0.8$(fix) & $2.6^{+0.9}_{-0.9}$ & $2.6^{+0.9}_{-0.9}$\\
\tableline
$BRi$-LBG & 4.0 & $23.3\leq i'<24.8$ &  103 & 1 & $0.8$(fix) & $6.0^{+3.2}_{-3.2}$ & $6.1^{+3.3}_{-3.2}$\\
$BRi$-LBG & 4.0 & $23.8\leq i'<25.3$ &  353 & 1 & $0.8$(fix) & $2.4^{+0.9}_{-0.9}$ & $2.4^{+1.0}_{-1.0}$\\
$BRi$-LBG & 4.0 & $24.3\leq i'<25.8$ &  724 & 1 & $0.8$(fix) & $1.3^{+0.5}_{-0.5}$ & $1.3^{+0.5}_{-0.5}$\\
$BRi$-LBG & 4.0 & $24.8\leq i'<26.3$ & 1335 & 1 & $0.8$(fix) & $1.2^{+0.3}_{-0.2}$ & $1.3^{+0.3}_{-0.2}$\\
\tableline
$BRi$-LBG & 4.0 & $-0.2<E(B-V)<0.2$\tablenotemark{a} &  253 & 1 & $0.8$(fix) & $2.9^{+1.3}_{-1.3}$ & $2.9^{+1.3}_{-1.3}$\\
$BRi$-LBG & 4.0 & $-0.1<E(B-V)<0.3$\tablenotemark{a} &  268 & 1 & $0.8$(fix) & $2.0^{+1.2}_{-1.2}$ & $2.0^{+1.2}_{-1.3}$\\
$BRi$-LBG & 4.0 & $ 0.0<E(B-V)<0.4$\tablenotemark{a} &  228 & 1 & $0.8$(fix) & $3.2^{+1.5}_{-1.5}$ & $3.2^{+1.5}_{-1.5}$\\
$BRi$-LBG & 4.0 & $ 0.1<E(B-V)<0.5$\tablenotemark{a} &  153 & 1 & $0.8$(fix) & $2.6^{+2.1}_{-2.2}$ & $2.7^{+2.2}_{-2.2}$\\
$BRi$-LBG & 4.0 & $ 0.2<E(B-V)<0.6$\tablenotemark{a} &   76 & 1 & $0.8$(fix) & $6.0^{+4.3}_{-4.3}$ & $6.1^{+4.4}_{-4.4}$\\
\enddata
\tablenotetext{\dagger}{
Selection criteria of the sub-sample. ``all'' means the whole sample
of LBGs.
}
\tablenotetext{\dagger\dagger}{
Number of the sample galaxies.
}
\tablenotetext{\ddag}{
Fraction of contamination in the sample obtained
in SDS V.
}
\tablenotetext{\ddag\ddag}{
Correlation amplitude, $A_\omega$, corrected for
contamination.
}
\tablenotetext{\P}{
Clustering measurements for LBGs with $L\gtrsim L^*$
used to calculate $r_0$ and $b_g$ plotted 
in Figure \ref{fig:z_r0_bias}.
Corresponding $r_0$ and $b_g$ values 
are summarized in Table \ref{tab:r0_lstar}.
}
%
%
\tablenotetext{a}{
These $E(B-V)$-dependent subsamples are constructed 
from 357 bright LBGs with $i'<25.3$.
We calculate $E(B-V)$ from $i'-z'$ color 
(section \ref{sec:spatial_segregation_lbg_dust}). 
The $E(B-V)$ range of each subsample corresponds to 
$-0.18<i'-z'<0.16$ (subsample of $-0.2<E(B-V)<0.2$),
$-0.10<i'-z'<0.24$ ($-0.1<E(B-V)<0.3$),
$-0.01<i'-z'<0.33$ ($0.0<E(B-V)<0.4$),
$0.07<i'-z'<0.41$ ($0.1<E(B-V)<0.5$), 
and $0.16<i'-z'<0.50$ ($0.2<E(B-V)<0.6$).
}
%
%

\tablecomments{
Clustering amplitudes, $A_\omega^{\rm raw}$ and $A_\omega$, are
in units of arcsec$^\beta$.
We regard $A_\omega$ as the best estimates.
}

\end{deluxetable}

\clearpage

\begin{deluxetable}{cccrrrr}
\tabletypesize{\scriptsize}
\tablecaption{Summary of the correlation length and 
the galaxy-dark matter bias.
\label{tab:r0_comb}}
\tablewidth{0pt}
\tablehead{
\colhead{Sample Name} &
\colhead{$\langle z \rangle$} & 
\colhead{Criteria\tablenotemark{\dagger}} & 
\colhead{$r_0^{\rm raw}$} &
\colhead{$b_g^{\rm raw}$} &
\colhead{$r_0$} &
\colhead{$b_g$}
}
\startdata
$BRi$-LBG & 4.0 & all ($i'\lesssim26$) & $4.0^{+0.2}_{-0.2}$ & $2.8^{+0.1}_{-0.1}$ & $4.1^{+0.2}_{-0.2}$ & $2.9^{+0.1}_{-0.1}$\\
$Viz$-LBG & 4.7 & all ($z'<25.8$) \tablenotemark{\P} & $4.2^{+1.0}_{-1.2}$ & $3.4^{+0.7}_{-0.9}$ & $5.9^{+1.3}_{-1.7}$ & $4.6^{+0.9}_{-1.2}$\\
\tableline
$BRi$-LBG & 4.0 & $i'<25.3$\tablenotemark{\P} & $5.1^{+0.9}_{-1.1}$ & $3.5^{+0.6}_{-0.7}$ & $5.1^{+1.0}_{-1.1}$ & $3.5^{+0.6}_{-0.7}$\\ 
\tableline
$BRi$-LBG & 4.0 & $23.3\leq i'<24.8$ & $7.9^{+2.1}_{-2.7}$ & $5.3^{+1.3}_{-1.7}$ & $7.9^{+2.1}_{-2.7}$ & $5.3^{+1.3}_{-1.7}$\\
$BRi$-LBG & 4.0 & $23.8\leq i'<25.3$ & $4.7^{+1.0}_{-1.2}$ & $3.3^{+0.6}_{-0.7}$ & $4.8^{+1.0}_{-1.2}$ & $3.4^{+0.6}_{-0.8}$\\
$BRi$-LBG & 4.0 & $24.3\leq i'<25.8$ & $3.4^{+0.6}_{-0.7}$ & $2.5^{+0.4}_{-0.5}$ & $3.5^{+0.7}_{-0.8}$ & $2.5^{+0.4}_{-0.5}$\\
$BRi$-LBG & 4.0 & $24.8\leq i'<26.3$ & $3.3^{+0.4}_{-0.4}$ & $2.4^{+0.2}_{-0.2}$ & $3.4^{+0.4}_{-0.4}$ & $2.4^{+0.2}_{-0.2}$\\
\tableline
$BRi$-LBG & 4.0 & $-0.2<E(B-V)<0.2$\tablenotemark{a} & $5.4^{+1.3}_{-1.6}$ & $3.7^{+0.8}_{-1.0}$ & $5.5^{+1.3}_{-1.6}$ & $3.8^{+0.8}_{-1.0}$\\
$BRi$-LBG & 4.0 & $-0.1<E(B-V)<0.3$\tablenotemark{a} & $4.4^{+1.3}_{-1.8}$ & $3.1^{+0.8}_{-1.2}$ & $4.5^{+1.4}_{-1.8}$ & $3.1^{+0.8}_{-1.2}$\\
$BRi$-LBG & 4.0 & $ 0.0<E(B-V)<0.4$\tablenotemark{a} & $5.7^{+1.3}_{-1.7}$ & $3.9^{+0.8}_{-1.0}$ & $5.8^{+1.4}_{-1.7}$ & $3.9^{+0.8}_{-1.1}$\\
$BRi$-LBG & 4.0 & $ 0.1<E(B-V)<0.5$\tablenotemark{a} & $5.1^{+2.0}_{-3.2}$ & $3.5^{+1.2}_{-2.1}$ & $5.2^{+2.0}_{-3.2}$ & $3.6^{+1.2}_{-2.1}$\\
$BRi$-LBG & 4.0 & $ 0.2<E(B-V)<0.6$\tablenotemark{a} & $8.1^{+2.8}_{-4.2}$ & $5.4^{+1.7}_{-2.6}$ & $8.2^{+2.9}_{-4.2}$ & $5.4^{+1.7}_{-2.6}$\\
\enddata
\tablenotetext{\dagger}{
Selection criteria of the sub-sample. ``all'' means the whole sample
of LBGs.
}
\tablenotetext{\P}{
Clustering measurements for LBGs with $L\gtrsim L^*$ 
plotted in Figure \ref{fig:z_r0_bias}.
These are repeatedly presented in Table \ref{tab:r0_lstar}.
}
%
%
\tablenotetext{a}{
These $E(B-V)$-dependent subsamples are constructed 
from 357 bright LBGs with $i'<25.3$.
We calculate $E(B-V)$ from $i'-z'$ color 
(section \ref{sec:spatial_segregation_lbg_dust}). 
The $E(B-V)$ range of each subsample corresponds to 
$-0.18<i'-z'<0.16$ (subsample of $-0.2<E(B-V)<0.2$),
$-0.10<i'-z'<0.24$ ($-0.1<E(B-V)<0.3$),
$-0.01<i'-z'<0.33$ ($0.0<E(B-V)<0.4$),
$0.07<i'-z'<0.41$ ($0.1<E(B-V)<0.5$), 
and $0.16<i'-z'<0.50$ ($0.2<E(B-V)<0.6$).
}
%
%

\tablecomments{
$r_0$ and $b_g$ are the correlation length and bias parameter,
respectively, calculated from the contamination-corrected clustering
amplitude, $A_\omega$, while $r_0^{\rm raw}$ and $b_g^{\rm raw}$ 
are those calculated
from the raw clustering amplitude, $A_\omega^{\rm raw}$ 
(see Table \ref{tab:acorr_comb}).
We regard $r_0$ and $b_g$ as for the best estimates.
$r_0$ and $r_0^{\rm raw}$
are in units of $h_{100}^{-1} $ Mpc.
$b_g$ and $b_g^{\rm raw}$
are defined at $8h_{100}^{-1}$Mpc.
}
\end{deluxetable}

\clearpage

\begin{deluxetable}{cccrr}
\tabletypesize{\scriptsize}
\tablecaption{Summary of the clustering measurements 
for LBGs with $L\gtrsim L^*$ and all LAEs 
plotted in Figure \ref{fig:z_r0_bias}.
\label{tab:r0_lstar}}
\tablewidth{0pt}
\tablehead{
\colhead{Sample Name} &
\colhead{$\langle z \rangle$} & 
\colhead{Criteria\tablenotemark{\P}} & 
\colhead{$r_0$\tablenotemark{\dagger}} &
\colhead{$b_g$\tablenotemark{\dagger}}
}
\startdata
$U_nGR$-LBG(SPEC)\tablenotemark{\ddag} & 3.0 & $M_{\rm 1700}<-20.5\simeq M^*$ ($R< 25.0$) & $5.0^{+0.7}_{-0.7}$ & $2.7^{+0.4}_{-0.4}$\\
$BRi$-LBG & 4.0 & $M_{\rm 1700}<-20.8\simeq M^*$  ($i'<25.3$) & $5.1^{+1.0}_{-1.1}$ & $3.5^{+0.6}_{-0.7}$\\ 
$Viz$-LBG & 4.7 & $M_{\rm 1700}<-20.5\simeq M^*$  ($z'<25.8$) & $5.9^{+1.3}_{-1.7}$ & $4.6^{+0.9}_{-1.2}$\\
\tableline
LAE\tablenotemark{\ddag\ddag}     & 4.9 &   $M_{\rm 1700}<-19.0$           ($z'<27.3$) & $6.2^{+0.5}_{-0.5}$ & $4.9^{+0.4}_{-0.4}$\\
\enddata
\tablenotetext{\dagger}{
$r_0$ and $b_g$ for $BRi$-LBGs, $Viz$-LBGs, and LAEs 
are contamination-corrected values.
}
\tablenotetext{\ddag}{
$r_0$ and $\gamma$ of $U$-band drop-out LBGs
are obtained from the SPEC sample of \citet{giavalisco2001}.
}
\tablenotetext{\ddag\ddag}{
$r_0$ and $b_g$
are obtained by SDS II.
These are contamination-corrected values.
}
\tablenotetext{\P}{
These criteria correspond to $L\gtrsim L^*$ for LBGs
at each redshift (see SDS V).
}
\tablecomments{
Values for $BRi$-LBGs and $Viz$-LBGs 
are the same as those presented in
Table \ref{tab:r0_comb}.
}

\end{deluxetable}

\clearpage

\begin{deluxetable}{cccccccc}
\tabletypesize{\scriptsize}
\tablecaption{Mass of hosting dark halos. 
\label{tab:halo_mass}}
\tablewidth{0pt}
\setlength{\tabcolsep}{0.02in}
\tablehead{
\colhead{Sample Name} &
\colhead{$\langle z \rangle$} & 
\colhead{Selection Criteria} & 
\colhead{$\langle M \rangle [h_{70}^{-1} M_\odot]$} &
\colhead{$\log(n)$} &
\colhead{$N_{\rm occup}$} &
\colhead{$\langle M \rangle (z=0) [h_{70}^{-1} M_\odot]$} &
\colhead{$N^{\rm min}$(merge)}\\
}
\startdata
$U_nGR$-LBG(PHOT)\tablenotemark{\ddag} & 3.0 & $R<25.5$      & $1.4^{+2.6}_{-1.1}\times 10^{11}$ & $-3.4\pm0.06$ & 0.04 & $2.8^{+1.6}_{-1.2}\times 10^{13}$ & $1.4$\\
$U_nGR$-LBG(SPEC)\tablenotemark{\ddag} & 3.0 & $R<25.0$\tablenotemark{\P} & $1.3^{+1.0}_{-0.6}\times 10^{12}$ & $-3.6\pm0.06$ & 0.3 & $7.1^{+2.9}_{-2.1}\times 10^{13}$ & $1.1$\\
$UBV$-LBG(HDF)\tablenotemark{\ddag}  & 2.6 & $V_{606}<27.0$& $<1 \times 10^{10}$               & $-2.7\pm0.06$ & \nodata & \nodata & \nodata \\
FIRES-All      & 3.0 & $K<24$        & $2.2^{+1.3}_{-1.0}\times 10^{12}$ & $-2.9\pm0.03$ & 3.5 & $1.0^{+0.3}_{-0.3}\times 10^{14}$ & $6.6$\\
FIRES-Red      & 3.0 & $K<24$ \& $J-K>1.7$ & $8.9^{+3.7}_{-3.9}\times 10^{12}$ & $-3.2\pm0.05$ & 16.0 & $2.0^{+0.5}_{-0.6}\times 10^{14}$ & $4.7$ \\
FIRES-Blue     & 3.0 & $K<24$ \& $J-K<1.7$ & $3.5^{+14}_{-3.6}\times 10^{11}$ & $-3.1\pm0.02$  & 0.3 & $4.0^{+4.9}_{-4.0}\times 10^{13}$  & $3.4$ \\
SCUBA          & $\sim3$ & $S_{\rm 850}>3$mJy  & $2.8^{+2.8}_{-1.9}\times 10^{13}$ & $-5.0\pm0.5$ & 3.5 & $4.5^{+2.6}_{-2.5}\times 10^{14}$ & $0.1$\\
\tableline
$BRi$-LBG & 4.0 & $i'\lesssim26$      & $4.1^{+1.0}_{-0.7}\times 10^{11}$ & $-3.2\pm0.03$ & 0.3 & $6.5^{+0.6}_{-0.5}\times 10^{13}$ & $2.8$\\
$BRi$-LBG & 4.0 & $i'<25.3$\tablenotemark{\P}      & $1.0^{+0.9}_{-0.7}\times 10^{12}$ & $-3.8\pm0.05$ & 0.3 & $9.6^{+3.4}_{-3.5}\times 10^{13}$ & $0.9$\\
$BRi$-LBG & 4.0 & $23.3<i'<24.8$ ($i'\simeq 24.0$) & $4.6^{+4.0}_{-3.4}\times 10^{12}$ & $-4.3\pm0.1$  & 1.3 & $2.1^{+1.2}_{-1.1}\times 10^{14}$ & $0.4$\\
$BRi$-LBG & 4.0 & $23.8<i'<25.3$ ($i'\simeq 24.5$) & $8.6^{+8.9}_{-6.2}\times 10^{11}$ & $-3.8\pm0.09$ & 0.3 & $8.9^{+3.5}_{-3.6}\times 10^{13}$ & $0.9$\\
$BRi$-LBG & 4.0 & $24.3<i'<25.8$ ($i'\simeq 25.0$) & $2.1^{+2.6}_{-1.5}\times 10^{11}$ & $-3.3\pm0.08$ & 0.1 & $5.0^{+1.9}_{-1.8}\times 10^{13}$ & $2.1$\\
$BRi$-LBG & 4.0 & $24.8<i'<26.3$ ($i'\simeq 25.5$) & $1.7^{+1.1}_{-0.7}\times 10^{11}$ & $-2.9\pm0.07$ & 0.2 & $4.6^{+0.1}_{-0.9}\times 10^{13}$ & $4.8$\\
\tableline
$Viz$-LBG & 4.7 & $z'<25.8$\tablenotemark{\P}       & $8.9^{+8.9}_{-6.4}\times 10^{11}$ & $-3.8\pm0.3$  & 0.6 & $1.3^{+0.5}_{-0.6}\times 10^{14}$ & $0.9$\\
LAE     & 4.9 & $NB711<25.8$      & $1.1^{+0.5}_{-0.2}\times 10^{12}$\tablenotemark{\dagger} & $-3.2\pm0.3$ & 3.2\tablenotemark{\dagger} & $1.4^{+0.2}_{-0.2}\times 10^{14}$ & $3.8$\\
\enddata
\tablenotetext{\ddag}{
$U$-band drop-out LBG samples given by \citet{giavalisco2001}.
SPEC, PHOT, and HDF mean the spectroscopic sample ($R<25$),
photometric sample ($R<25.5$), and photometric sample of HDF-N ($R<27$),
respectively.
}
\tablenotetext{\dagger}{Maximum mass and occupation number of 
LAEs are $3.5^{+1.4}_{-0.7}\times 10^{12}$ $h_{70}^{-1} M_\odot$ 
and $35$ (see footnote 
in section \ref{sec:galaxy_hosting_z5}).
}
\tablenotetext{\P}{
Samples of LBGs with $L\gtrsim L^*$ at $z=3$, $4$, and $5$.
}
\tablecomments{
$\langle M \rangle [h_{70}^{-1} M_\odot]$ and 
$\langle M \rangle (z=0) [h_{70}^{-1} M_\odot]$
are the mass of 
dark halos hosting high-$z$ galaxies and the typical mass of 
their descendants at $z=0$, respectively.
$\log(n)$ is the number density of high-$z$ galaxies
in units of mag$^{-1}$ $h_{70}^{3}$ Mpc$^{-3}$.
$N_{\rm occup}$ is the occupation number, and
$N^{\rm min}$(merge) is the minimum number of mergers
required from high-$z$ to $z=0$, 
assuming that the high-$z$ galaxies evolve into present-day
galaxies with $M_{b_g}\lesssim -19$ mag.
}

\end{deluxetable}

\clearpage

\begin{deluxetable}{cccccccc}
\tabletypesize{\scriptsize}
\tablecaption{Mass of hosting dark halos for LBGs with $L\gtrsim L^*$
and all LAEs. 
\label{tab:halo_mass_lstar}}
\tablewidth{0pt}
\setlength{\tabcolsep}{0.02in}
\tablehead{
\colhead{Sample Name} &
\colhead{$\langle z \rangle$} & 
\colhead{Selection Criteria} & 
\colhead{$\langle M \rangle [h_{70}^{-1} M_\odot]$} &
\colhead{$\log(n)$} &
\colhead{$N_{\rm occup}$} &
\colhead{$\langle M \rangle (z=0) [h_{70}^{-1} M_\odot]$} &
\colhead{$N^{\rm min}$(merge)}\\
}
\startdata
$U_nGR$-LBG(SPEC)\tablenotemark{\ddag} & 3.0 & $R<25.0$ & $1.3^{+1.0}_{-0.6}\times 10^{12}$ & $-3.6\pm0.06$ & 0.3 & $7.1^{+2.9}_{-2.1}\times 10^{13}$ & $1.1$\\
$BRi$-LBG & 4.0 & $i'<25.3$     & $1.0^{+0.9}_{-0.7}\times 10^{12}$ & $-3.8\pm0.05$ & 0.3 & $9.6^{+3.4}_{-3.5}\times 10^{13}$ & $0.9$\\
$Viz$-LBG & 4.7 & $z'<25.8$       & $8.9^{+8.9}_{-6.4}\times 10^{11}$ & $-3.8\pm0.3$  & 0.6 & $1.3^{+0.5}_{-0.6}\times 10^{14}$ & $0.9$\\
\tableline
LAE     & 4.9 & $NB711<25.8$      & $1.1^{+0.5}_{-0.2}\times 10^{12}$ & $-3.2\pm0.3$ & 3.2 & $1.4^{+0.2}_{-0.2}\times 10^{14}$ & $3.8$\\
\enddata
\tablenotetext{\ddag}{
$U$-band drop-out LBGs given by \citet{giavalisco2001}.
}
\tablecomments{
$\langle M \rangle [h_{70}^{-1} M_\odot]$ and 
$\langle M \rangle (z=0) [h_{70}^{-1} M_\odot]$
are the mass of 
dark halos hosting high-$z$ galaxies and the typical mass of 
their descendants at $z=0$, respectively.
$\log(n)$ is the number density of high-$z$ galaxies
in units of mag$^{-1}$ $h_{70}^{3}$ Mpc$^{-3}$.
$N_{\rm occup}$ is the occupation number, and
$N^{\rm min}$(merge) is the minimum number of mergers
required from high-$z$ to $z=0$, 
assuming that the high-$z$ galaxies evolve into present-day
galaxies with $M_{b_g}\lesssim -19$ mag.
}

\end{deluxetable}

\clearpage



\end{document}